\documentclass[aps,pre,twocolumn,superscriptaddress,longbibliography,nobibnotes,nodoi,nourl,noeprint]{revtex4-2}
\usepackage{amsmath,amssymb,physics,bm,siunitx}
\usepackage{xcolor,graphicx,soul}
\usepackage[colorlinks=true,citecolor=blue,urlcolor=blue,linkcolor=black]{hyperref}
\usepackage{subfiles}

\usepackage{upgreek}
\newcommand{\rms}{\rm\scriptscriptstyle}

\newcommand{\kb}{k_{\rms B}}
\newcommand{\tepsA}{\tilde\epsilon^{\rm A}}
\newcommand{\tepsB}{\tilde\epsilon^{\rm B}}

\newcommand{\epsA}{\epsilon_{\rm A}}
\newcommand{\epsB}{\epsilon_{\rm B}}

\newcommand{\mAA}{\bar\epsilon^{\rm A}_{\rm A}}
\newcommand{\mAB}{\bar\epsilon^{\rm A}_{\rm B}}
\newcommand{\mBA}{\bar\epsilon^{\rm B}_{\rm A}}
\newcommand{\mBB}{\bar\epsilon^{\rm B}_{\rm B}}

\begin{document}

\title{Internal friction and frequency-dependent conductivity in the bivariate site energy model
for the mixed alkali effect in glasses}

\author{Justus Leiber}
\email{jleiber@uni-osnabrueck.de}
\affiliation{Universit\"{a}t Osnabr\"uck, Institut f\"ur Physik, Barbarastra{\ss}e 7, D-49076 Osnabr\"uck, Germany}

\author{Sven Lohmann}
\email{slohmann@uni-osnabrueck.de}
\affiliation{Universit\"{a}t Osnabr\"uck, Institut f\"ur Physik, Barbarastra{\ss}e 7, D-49076 Osnabr\"uck, Germany}

\author{Quinn Emilia Fischer}
\email{qfischer@uni-osnabrueck.de}
\affiliation{Universit\"{a}t Osnabr\"uck, Institut f\"ur Physik, Barbarastra{\ss}e 7, D-49076 Osnabr\"uck, Germany}

\author{Philipp Maass} 
\email{maass@uni-osnabrueck.de}
\affiliation{Universit\"{a}t Osnabr\"uck, Institut f\"ur Physik, Barbarastra{\ss}e 7, D-49076 Osnabr\"uck, Germany}

\date{November 2, 2026}

\begin{abstract}
The mixed alkali effect is fundamental to understanding ion transport in glasses. 
It manifests itself in strong changes of various ionic transport properties when one type of ion is successively replaced by another type. Recently, we have developed
a new theory of thermally activated hopping in bivariate random site energy landscapes
that takes into account site occupation statistics according to generalized Fermi distributions and mixture kinetics in accordance with linear irreversible thermodynamics. 
The theory was successfully applied to predict the behavior of partial conductivities and their activation energies at low frequencies corresponding to the dc limit. Here we show that the theory is capable to capture the salient features of the mixed alkali effect observed in ac ion conductivities and in the peculiar behavior of the internal friction.
\end{abstract}

\maketitle

\section{Introduction}
The mixed alkali effect (MAE), or mixed mobile ion effect, is a universal phenomenon observed in ion-conducting glasses when two or more types of mobile ions are mixed \cite{Isard:1969, Day:1976, Ingram:1987, Dyre/etal:2009}. While it is commonly studied by mixing alkali ions, this effect has also been observed in glasses containing mobile silver and thallium ions \cite{Bartholomew:1973, Carette/etal:1983, Machida/etal:1995, Rau/etal:2001, Bokova/etal:2015, Bokova/etal:2020}, protons \cite{Molinelli/etal:1985, Garcia/etal:2002}, and by replacing fluorine with chlorine ions \cite{Almeida/Mackenzie:1982, Kahnt/Reau:1990, Alenezi/etal:2026}.
The MAE manifests itself through strong variations 
of ion transport properties 
when one type A of mobile ion type is substituted by another type B.
Most striking is the behavior of the activation energies $E_{\rm A}$ and $E_{\rm B}$
of the ion mobilities: when A ions are replaced by B ions, $E_{\rm A}$ increases, while
$E_{\rm B}$ decreases.
At low temperatures, this results in mobility changes
spanning several orders of magnitude.
As a function of the mixing ratio, the dc partial 
conductivities $\sigma^{\rm dc}_{\rm A}$ and $\sigma^{\rm dc}_{\rm B}$ of the two ion types generally cross. 
Close to the crossing ratio, the total ion 
conductivity exhibits a deep minimum. This minimum can be thousands of times lower
than the ion conductivities of the respective single-ionic glasses.

Theoretically, it has been proposed that mobile ions of different types 
prefer distinct structural environments within the glassy network \cite{Maass/etal:1992}
and possess different site energies \cite{Hunt:1999, Baranovskii/Cordes:1999, Maass:1999, Kirchheim:2000}.
The origin of the MAE can be understood through the difference in site energies between A and B ions.
EXAFS \cite{Greaves:1998}
and NMR studies \cite{Tsuchida/etal:2012, NogueiradosSantos/etal:2025} have demonstrated
the structural specificity of sites preferentially occupied by mobile ion species.
The MAE is also studied in
molecular dynamics simulations 
\cite{Habasaki/etal:1995, Habasaki/etal:2004,
  Balbuena/etal:2014, Lodesani/etal:2020, Noritake/Naito:2023}. Specifically, simulations
of mixed alkali silicate glasses have highlighted
the importance of distinct site energy landscapes \cite{Lammert/Heuer:2005}. 
 
Recently, we have developed a theory of the MAE that consistently incorporates the statistical
mechanics and kinetics of the mobile ion mixture \cite{Leiber/etal:2026}. 
This theory, which we term
the bivariate site energy model (BSEM), considers a joint probability density of site
energy states.
The mean occupation numbers of these states are given by generalized Fermi distributions 
that depend on the chemical potentials of both ion
types. Moreover, we evaluated the off‑diagonal Onsager coefficients in the modeling, which
describe the cross-terms in the linear response of ionic currents to driving fields.

Monte Carlo simulations of the BSEM yielded
variations of the $\sigma^{\rm dc}_\upalpha$
and their activation energies $E_\upalpha$ with the mixing ratio as typically
observed in experiments ($\upalpha={\rm A, B}$). By generalizing a theory for
many-particle hopping conduction 
\cite{Miller/Abrahams:1960, Ambegaokar/etal:1971} to a two-component
charge carrier mixture, we developed an efficient method for approximate numerical calculations of
$\sigma^{\rm dc}_\upalpha$ and $E_\upalpha$. The results of these calculations
showed good agreement 
with the Monte Carlo simulations. The numerical method allows for the determination of
$\sigma^{\rm dc}_\upalpha$ and $E_\upalpha$ in mere minutes on a standard personal computer. 

A systematic analysis of the BSEM parameters revealed
that convex or concave behaviors of $\sigma^{\rm dc}_\upalpha$ and $E_\upalpha$ 
as a function of the mixing ratio can be
related to different types of onsite correlations between A and B energies. 
The strong vulnerability observed in some mixed alkali glasses,
which amounts to a stronger than exponential
decrease of the mobility of the majority ion upon initial replacement by the minority ion,
can be reproduced when including spatial correlations between the site energies, as suggested previously \cite{Peibst/etal:2005}. Furthermore,
the fast numerical method enables the
fitting of BSEM parameters for a quantitative description of experimental findings. 
In Ref.~\cite{Leiber/etal:2026},
we demonstrated this for various mixed alkali phosphate glasses.

Thus far, our focus has been on predicting dc ionic conductivities. 
In this study, we
investigate whether the BSEM is capable to capture further striking features of the MAE: the weaking of the effect with frequency when considering ac conductivities, and the peculiar behavior of the internal friction observed in
mechanical relaxation experiments \cite{Shelby:1965}.

\section{Bivariate site energy model (BSEM)}
\label{sec:BSEM}
This section introduces the BSEM and the following Sec.~\ref{sec:partial-conductivities}
shows how partial ionic conductivities
are derived. A more detailed description is given in Ref.~\cite{Leiber/etal:2026}.

\subsection{Hopping transport}
\label{subsec:hopping-transport}
In the BSEM, 
each site $i$ in the glassy network can be occupied by at most one ion of type A or B, or by a
vacancy V. The fractions of occupied sites are
\begin{equation}
c_\upalpha=\frac{N_\upalpha}{M}\,,
\end{equation}
where $N_\upalpha$ is the number of $\upalpha$ ions and
 $M$ the total number of sites.
The mixing ratio is
\begin{equation}
x=\frac{c_{\rm A}}{c_{\rm A}+c_{\rm B}}\,.
\end{equation}
The total energy is
\begin{equation}
H=H_0+\sum_{i=1}^M (\epsilon_i^{\rm A} n_i^{\rm A}+\epsilon_i^{\rm B} n_i^{\rm B})\,,
\label{eq:H2}
\end{equation}
where $H_0$ is the energy if all sites are vacant,
$\epsilon_i^\upalpha$ are the site energies of $\upalpha$ ions, and
\begin{equation}
n^\upalpha_i=\left\{\begin{array}{cc}
1, & \mbox{if site $i$ is occupied by an $\upalpha$ ion}, \\[0.5ex]
0, & \mbox{otherwise},
\end{array}\right.
\end{equation}
are occupation numbers.

\begin{figure}[t!]
\centering
\includegraphics[width=0.4\textwidth]{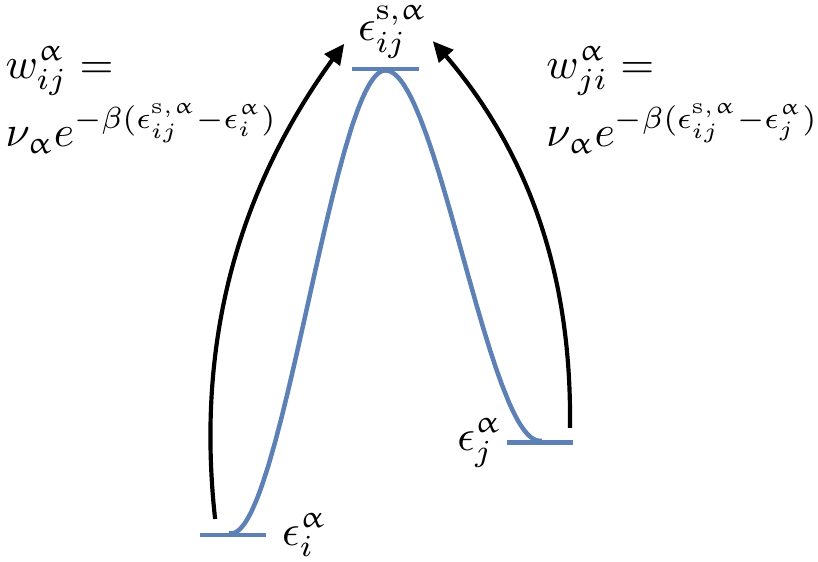}
\caption{\small Possible thermally activated
jumps of an $\upalpha$ ion, $\upalpha={\rm A,B}$, between neighboring sites
$i$ and $j$ with site energies $\epsilon_i^\upalpha$ and $\epsilon_j^\upalpha$ in the glassy network.
The $\upalpha$ ion has to surmount a saddle point with energy $\epsilon_{ij}^{{\rm s}, \upalpha}$, leading to
the jump
rates $w_{ij}^\upalpha=\nu_\upalpha e^{-\beta(\epsilon_{ij}^{\rm s}-\epsilon_i)}$
and $w_{ji}^\upalpha=\nu_\upalpha e^{-\beta(\epsilon_{ij}^{\rm s}-\epsilon_j)}$
according to
chemical kinetics.}
\label{fig:illustration_jump_rates}
\end{figure}

A and B ions can jump to vacant nearest-neighbor sites. 
As illustrated in Fig.~\ref{fig:illustration_jump_rates},
the jump rate of an $\upalpha$ ion occupying
site $i$ to a vacant nearest-neighbor site $j$ is
\begin{equation}
w_{ij}^{0,\upalpha}=
\nu_\upalpha e^{-\beta(\epsilon_{ij}^{{\rm s},\upalpha}-\epsilon^\upalpha_i)}\,.
\label{eq:wij0}
\end{equation}
Here, $\nu_\upalpha$ is an attempt frequency, $\beta=1/\kb T$ is the
inverse thermal energy, 
and $\epsilon_{ij}^{{\rm s},\upalpha}=\epsilon_{ji}^{{\rm s},\upalpha}$ is the saddle point
energy.

When applying a small external electric field ${\vb E}$, the jump
rates are
\begin{equation}
w_{ij}^{\upalpha}=w_{ij}^{0,\upalpha}\,\left(1+\frac{B_{ij}^\upalpha}{2}\right)\,,
\label{eq:w}
\end{equation}
where
\begin{equation}
B_{ij}^\upalpha=\beta q_\upalpha\, \vb E\cdot \vb R_{ij}=-B_{ji}^\upalpha\,,
\label{eq:B}
\end{equation}
$q_\upalpha$ is the charge of ions of type $\upalpha$, and $\vb
R_{ij}=(\vb R_j\!-\!\vb R_i)$ is the vector pointing from position
$\vb R_i$ of site $i$ to position $\vb R_j$ of site $j$.

\subsection{Site energy distribution}
\label{subsec:site-energy-distribution}
At a site $i$, the A and B energies assume values in the two-dimensional interval 
$(\epsilon_{\rm A},\epsilon_{\rm A}+\dd\epsilon_{\rm A})\times
(\epsilon_{\rm B},\epsilon_{\rm B}+\dd\epsilon_{\rm B})$ 
with a probability $\psi(\epsilon_{\rm A},\epsilon_{\rm B})\dd \epsilon_{\rm A}\dd \epsilon_{\rm B}$
given by a bivariate
density $\psi(\epsilon_{\rm A},\epsilon_{\rm B})$.
We assume it to be of the form
\begin{equation}
\psi(\epsilon_{\rm A}, \epsilon_{\rm B})
=xG_{\rm A}(\epsilon_{\rm A}, \epsilon_{\rm B})
+(1-x)G_{\rm B}(\epsilon_{\rm A}, \epsilon_{\rm B})\,.
\label{eq:joint-PDF}
\end{equation}
This form is motivated by the fact that in
melt-grown mixed alkali glasses, the glassy network tends to provide  
structural environments favorable for A and B ions with probabilities $x$ and $1-x$.
Favorable sites for A ions have $\epsilon_i^{\rm A}<\epsilon_i^{\rm B}$, i.e.\ the
$G_{\rm A}(\epsilon_{\rm A}, \epsilon_{\rm B})$ should be centered
around energies $(\mAA,\mAB)$ left of the diagonal $\epsA=\epsB$.
Analogously, $G_{\rm B}(\epsilon_{\rm A}, \epsilon_{\rm B})$ is centered
around energies $(\mBA,\mBB)$ right of the diagonal $\epsA=\epsB$.

A site energy of an $\upalpha$ ion predominantly results from local interactions
with atoms of the surrounding host network, i.e.\ many additive contributions with
random spatial fluctuations.
We hence assume the 
$G_\upalpha(\epsilon_{\rm A}, \epsilon_{\rm B})$ to be Gaussian according to the central limit theorem:
\begin{align}
&G_\upalpha(\epsA,\epsB)=\frac{1}{2\pi\sqrt{\det C^\upalpha}}
\label{eq:Galpha}\\
&\hspace{3em}{}\times\exp[-\frac{1}{2}
\!\sum_{\upbeta,\upgamma={\rm A,B}}\!(\epsilon_\upbeta\!-\!\bar\epsilon^\upalpha_\upbeta) 
(C^\upalpha)^{-1}_{\upbeta\upgamma}(\epsilon_\upgamma\!-\!\bar\epsilon^\upalpha_\upgamma)]\,.
\nonumber
\end{align}
The $(C^\upalpha)^{-1}_{\upbeta\upgamma}$ are the elements of the inverse covariance matrices
\begin{equation}
C^\upalpha=
\begin{pmatrix}
C^\upalpha_{\rm AA} & C^\upalpha_{\rm AB}\\[0.5ex]
C^\upalpha_{\rm AB} & C^\upalpha_{\rm BB}
\end{pmatrix}
=
\begin{pmatrix}
\Delta_{\rm A}^{\upalpha\,2} & \Delta_{\rm A}^\upalpha\Delta_{\rm B}^\upalpha\rho_\upalpha\\[0.5ex]
\Delta_{\rm A}^\upalpha\Delta_{\rm B}^\upalpha\rho_\upalpha & \Delta_{\rm B}^{\upalpha\,2} 
\end{pmatrix}\,.
\end{equation}
The parameters have the following physical significance: 
\newcounter{listnum}
\begin{list}{\alph{listnum})}{\setlength{\leftmargin}{1.5em}\setlength{\rightmargin}{0em}
\setlength{\itemsep}{0ex}\setlength{\topsep}{0ex}\usecounter{listnum}}

\item $\bar\epsilon^\upalpha_\upbeta$ is the mean energy of $\upbeta$ ions on sites 
favorably adapted to $\upalpha$ ions.

\item $\Delta^\upalpha_\upbeta$ quantifies how strongly the site energy of $\upbeta$ ions fluctuates
on sites favorably adapted to $\upalpha$ ions.

\item The Pearson correlation coefficient $\rho_\upalpha\in(-1,1)$ specifies how A and B energies on a
site favorably adapted to $\upalpha$ ions are correlated: $\rho_\upalpha>0$ ($\rho_\upalpha<0$) implies
a positive (negative) correlation, i.e.\ an A energy higher than the mean likely entails
a B energy higher (lower) than the mean.

\end{list}
The difference between $\bar\epsilon^\upalpha_\upbeta$ and $\bar\epsilon^\upalpha_\upalpha$
defines a mismatch
energy
\begin{equation}
\epsilon_{\rm mis}^\upbeta=\bar\epsilon^\upalpha_\upbeta-\bar\epsilon^\upalpha_\upalpha\,,
\end{equation}
which gives the mean additional energy if an $\upalpha$-adapted site is occupied by a $\upbeta$ rather 
than an $\upalpha$ ion.

\begin{figure}[t!]
\centering
\includegraphics[width=\columnwidth]{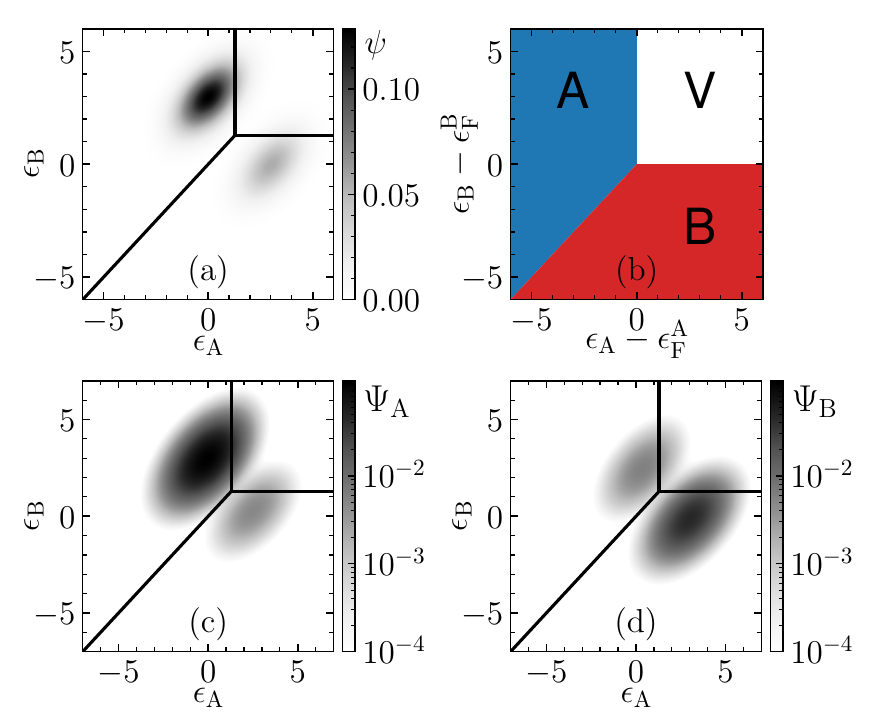}
\caption{\small (a) Bivariate site energy distribution [Eqs.~\eqref{eq:joint-PDF}, \eqref{eq:Galpha}], for a mixing ratio $x=0.7$, vacancy concentration
$c_{\rm V}=0.1$, and a symmetric set of parameters describing the energy landscape:
$\Delta_{\rm A}^{\rm A}=\Delta_{\rm B}^{\rm A}
=\Delta_{\rm B}^{\rm B}=\Delta_{\rm A}^{\rm B}=1$ (energy unit), $\rho_{\rm A}=\rho_{\rm B}=0.5$, 
$\bar\epsilon^{\rm A}_{\rm A}=\bar\epsilon^{\rm B}_{\rm B}=0$, and 
$\bar\epsilon^{\rm A}_{\rm B}=\bar\epsilon^{\rm B}_{\rm A}=3$. (b) Occupation of sites by A, B ions and vacancies V in the low-temperature limit
$T\to0$ in the $\epsilon_{\rm A}\epsilon_{\rm B}$-plane, where site energies are taken relative to the Fermi energies $\epsilon_{\rm F}^{\rm A}$ and $\epsilon_{\rm F}^{\rm B}$. For finite $T$, the Fermi edges given by $\epsilon_{\rm A}-\epsilon_{\rm F}^{\rm A}=0$ 
and $\epsilon_{\rm B}-\epsilon_{\rm F}^{\rm B}=0$
are smeared out with a width of order $\kb T$,
as well as the diagonal separating the regions of A and B occupation.
(c) Probability density $\Psi_{\rm A}(\epsilon_{\rm A},\epsilon_{\rm B},T)=\psi(\epsilon_{\rm A},\epsilon_{\rm B}) f(\beta\tilde\epsilon_{\rm A},\beta\tilde\epsilon_{\rm B})$ 
of finding a site with energies $(\epsilon_{\rm A},\epsilon_{\rm B})$ to be occupied 
by an A ion for the bivariate
site energy distribution in (a) at a thermal energy $\kb T=1/\beta=\Delta^{\rm A}_{\rm A}$.  
(d) Same as (c) for B ions.}
\label{fig:sed-fermi}
\end{figure}

An example of the bivariate site energy distribution \eqref{eq:joint-PDF} with the Gaussians
$G_\upalpha(\epsilon_{\rm A},\epsilon_{\rm B})$ from Eq.~\eqref{eq:Galpha}
is shown in Fig.~\ref{fig:sed-fermi}(a) for a mixing ratio $x=0.7$.

\subsection{Two-species Fermi distributions}
The probability of a site $i$ with energies $(\epsilon_i^{\rm A},\epsilon_i^{\rm B})$ 
to be occupied by an $\upalpha$ ion is given by the mean occupation number $\langle n_i^\upalpha\rangle$.  In equilibrium, the mean occupation numbers are
\begin{subequations}
\label{eq:meanoccupation-two-species}
\begin{align}
\langle n_i^{\rm A}\rangle_{\rm eq}
&=f_{\rm A}(\beta\tilde\epsilon_i^{\rm A},\beta\tilde\epsilon_i^{\rm B})=
f(\beta\tilde\epsilon_i^{\rm A},\beta\tilde\epsilon_i^{\rm B})\,,
\label{eq:meanA-two-species}\\[0.5ex]
\langle n_i^{\rm B}\rangle_{\rm eq}
&=f_{\rm B}(\beta\tilde\epsilon_i^{\rm A},\beta\tilde\epsilon_i^{\rm B})=
f(\beta\tilde\epsilon_i^{\rm B},\beta\tilde\epsilon_i^{\rm A})\,,
\label{eq:meanB-two-species}
\end{align}
\end{subequations}
where
\begin{equation}
\tilde\epsilon_i^\upalpha=\epsilon_i^\upalpha-\mu_\upalpha\,,
\label{eq:tildeeps}
\end{equation}
$\mu_\upalpha=\mu_\upalpha(T, c_{\rm A}, c_{\rm B})$ are the chemical potentials,
and
\begin{equation}
f(x,y)=\frac{1}{e^x+e^{x-y}+1}
\label{eq:two-species-fermi-function}
\end{equation}
is the two-species Fermi function.
The determining equations for the chemical potentials in Eq.~\eqref{eq:tildeeps} are
\begin{subequations}
\label{eq:mu-determ}
\begin{align}
\hspace*{-2em}\int\!\dd\epsilon_{\rm A}\! \int\!\dd\epsilon_{\rm B}\, 
\psi(\epsilon_{\rm A}, \epsilon_{\rm B})\,
f(\beta(\epsilon_{\rm A}-\mu_{\rm A}), \beta(\epsilon_{\rm B}-\mu_{\rm B}))&=
c_{\rm A}\,,
\label{eq:mu-determ-a}\\
\hspace*{-2em}\int\!\dd\epsilon_{\rm A}\! \int\!\dd\epsilon_{\rm B}\, 
\psi(\epsilon_{\rm A}, \epsilon_{\rm B})\,
f(\beta(\epsilon_{\rm B}-\mu_{\rm B}), \beta(\epsilon_{\rm A}-\mu_{\rm A}))&=
c_{\rm B}\,.
\label{eq:mu-determ-b}
\end{align}
\end{subequations}
The chemical potentials approach the Fermi
energies
\begin{equation}
\epsilon_{\rm F}^\upalpha(c_{\rm A}, c_{\rm B})
=\lim_{T\to0} \mu_\upalpha(T, c_{\rm A}, c_{\rm B})\,.
\end{equation}
for $T\to0$, where the 
mean occupation numbers for the A ions become
\begin{equation}
\lim\limits_{T\to0} \langle n_i^{\rm A}\rangle_{\rm eq}=\left\{\begin{array}{cc}
1\,, & \epsilon_i^{\rm A}<\epsilon_{\rms F}^{\rm A}\mbox{ and } 
(\epsilon_{\rms F}^{\rm A}-\epsilon_i^{\rm A})>(\epsilon_{\rms F}^{\rm B}-\epsilon_i^{\rm B})\,,\\[1ex]
0\,, &  \epsilon_i^{\rm A}>\epsilon^{\rm A}_{\rms F}\mbox{ or } 
(\epsilon_{\rms F}^{\rm A}-\epsilon_i^{\rm A})<(\epsilon_{\rms F}^{\rm B}-\epsilon_i^{\rm B})\,,
\end{array}\right.
\label{eq:nAeqTzero-1}
\end{equation}
and the analogous equation for B ions follows by interchanging A with B. 

The occupation in the $T\to0$ limit given by the two-species Fermi distribution in the
$\tilde\epsilon_{\rm A}\tilde\epsilon_{\rm B}$-plane 
is displayed in Fig.~\ref{fig:sed-fermi}(b). It has Fermi edges at $\epsilon_{\rms F}^{\rm A}$ and $\epsilon_{\rms F}^{\rm B}$ that separate areas of A and B occupied energy states from empty ones. The
diagonal $\tilde\epsilon_{\rm A}=\tilde\epsilon_{\rm B}$
 in Fig.~\ref{fig:sed-fermi}(b) is the separating line between the energy states of A and B occupation.
At finite low $T$, the Fermi edges as well as the A-B separating line are smeared out
over an interval of order $\kb T$.

The two-species Fermi distribution $f(\beta\tilde\epsilon_\upalpha,\beta\tilde\epsilon_\upbeta)$
gives the probability of occupation of a site by an $\upalpha$ ion 
under the condition that the site
has energies $(\epsilon_{\rm A},\epsilon_{\rm B})$. The probability of finding a
site with energies in the two-dimensional interval $(\epsilon_{\rm A},\epsilon_{\rm A}+\dd\epsilon_{\rm A})\times
(\epsilon_{\rm B},\epsilon_{\rm B}+\dd\epsilon_{\rm B})$ to be
 occupied with an $\upalpha$ ion is $\psi(\epsilon_{\rm A},\epsilon_{\rm B}) f(\beta\tilde\epsilon_\upalpha,\beta\tilde\epsilon_\upbeta)\,\dd\epsilon_{\rm A}\dd\epsilon_{\rm B}$. As an example,
we show in Figs.~\ref{fig:sed-fermi}(c) and \ref{fig:sed-fermi}(d) the respective
probability densities $\Psi_{\rm A}(\epsilon_{\rm A},\epsilon_{\rm B},T)=
\psi(\epsilon_{\rm A},\epsilon_{\rm B}) f(\beta\tilde\epsilon_{\rm A},\beta\tilde\epsilon_{\rm B})
$ and
$\Psi_{\rm B}(\epsilon_{\rm A},\epsilon_{\rm B},T)=\psi(\epsilon_{\rm A},\epsilon_{\rm B}) f(\beta\tilde\epsilon_{\rm B},\beta\tilde\epsilon_{\rm A})$
for the bivariate site energy distribution in Fig.~\ref{fig:sed-fermi}(a) at a low temperature.
 
\begin{figure*}[t!]
\centering
\includegraphics[width=\textwidth]{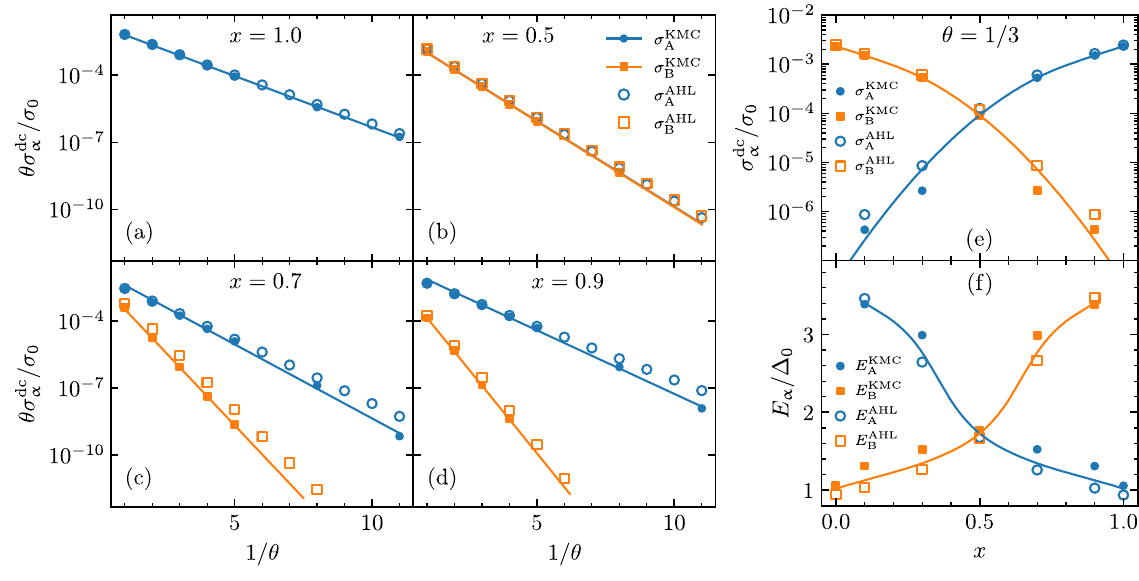}
\caption{\small (a)-(d) Arrhenius plots of partial conductivities at different mixing ratios. (e) Partial conductivities as a function of the mixing ratio $x$ at the scaled temperature $\theta=1/3$. (f) Activation energies of the partial conductivities
as a function of $x$. Results are shown for the vacancy concentration $c_{\rm V}=0.1$
and the symmetric parameter setting (see text) with $\epsilon_{\rm mis}=3$. 
Full and open symbols mark results from KMC simulations and
the generalized AHL theory, respectively. The solid lines are drawn as guide to the eye. 
}
\label{fig:partial-conductivities}
\end{figure*}

\section{Partial conductivities}
\label{sec:partial-conductivities}
Complex partial conductivities $\hat\sigma_\upalpha(\omega,T)$ can be accurately determined
by kinetic Monte Carlo (KMC) simulations in the presence of an electric field \cite{Maass/etal:1995, Lohmann/etal:2026}. An approximate method for their calculation
is provided by mapping the many-ion dynamics
with site exclusion interaction in the BSEM onto a dynamics of noninteracting ions in a disordered barrier landscape, where all sites have the same energy \cite{Leiber/etal:2026}.
In this mapping, correlations between occupation numbers in nonequilibrium states 
are neglected. The approximate calculation yields quite good agreement with results
for partial conductivities obtained from KMC simulations
in wide parameter regimes. 

For each ion type $\upalpha$, an individual barrier landscape
is obtained, with barriers of heights
\begin{equation}
u_{ij}^\upalpha = \epsilon_{ij}^{\rm s,\upalpha}
- \mu_\upalpha - \frac{1}{\beta}\ln\left[v(\beta\tepsA_i, \beta\tepsB_i)\right]
- \frac{1}{\beta}\ln\left[v(\beta\tepsA_j, \beta\tepsB_j)\right]\,,
\label{eq:uija}
\end{equation}
where 
\begin{align}
v(x,y)=e^xf(x,y)=e^yf(y,x)=\frac{1}{1+e^{-x}+e^{-y}}
\label{eq:v-f}
\end{align}
is the vacancy Fermi function: $v(\beta\tepsA_i, \beta\tepsB_i)$ gives the
probability of the site $i$ with energies $(\epsilon_i^{\rm A},\epsilon_i^{\rm B})$ to be vacant.

The barriers in Eq.~\eqref{eq:uija} change with temperature and they depend on the ionic concentrations  $c_\upalpha$ via the chemical potentials $\mu_\upalpha=\mu_\upalpha(T,c_{\rm A},c_{\rm B})$. For calculating 
frequency-dependent $\hat\sigma_\upalpha(\omega,T)$ of noninteracting particles in the barrier landscape, a computationally efficient numerical procedure is the velocity autocorrelation
(VAC) method \cite{Schroeder:2000, Schroeder:2008}.
For dc values
\begin{equation}
\sigma_\upalpha^{\rm dc}(T)=\lim\limits_{\omega\to0}\hat\sigma_\upalpha(\omega,T)\,,
\end{equation}
a network of link conductances can be considered.
An alternative method for calculating $\sigma^{\rm dc}_\upalpha(T)$ then
is to solve Kirchoff's equations with an iterative Kron reduction \cite{Bosi/Maass:2022}.
Both the VAC method and the numerical solution of Kirchoff's equations
are applicable even at very low $T$, where KMC simulations would need astronomically long computing times. 
If one is interested in only the activation energies $E_\upalpha$ of the $\sigma^{\rm dc}_\upalpha(T)$
without extracting them from Arrhenius plots, percolation theory can be employed 
\cite{Bosi/etal:2021}.

For detailed calculations of transport properties in the BSEM, one would need 
information about the location of the ionic sites in the glassy network. Such information may be obtained
from MD simulations and reverse Monte Carlo modeling \cite{Schuch/etal:2012, Karlsson/etal:2015}. We choose a simple cubic lattice with spacing $a$ here, which
does not represent an irregular set of sites in a glassy network.
However, due to the disorder in the site energy landscape, migration paths of the mobile ions are highly irregular also in the lattice. Hence, we do not expect the choice of the regular lattice 
to strongly affect the ionic transport behavior.
The lattice spacing $a$ is considered to represent a mean jump distance in a glassy network.

Figure~\ref{fig:partial-conductivities} shows an example of the behavior of dc partial conductivities and their activation energies. In this example, we have chosen
a vacancy concentration
$c_{\rm V}=0.1$, monovalent mobile ions with charge $q_\upalpha=e$ and a symmetric situation 
for the site energy distribution with only two energy parameters
\begin{equation}
\Delta_0\equiv\Delta_{\rm A}^{\rm A}=\Delta_{\rm B}^{\rm A}
=\Delta_{\rm B}^{\rm B}=\Delta_{\rm A}^{\rm B}\,,
\end{equation}
and
\begin{equation}
\epsilon_{\rm mis}\equiv\epsilon_{\rm mis}^{\rm A}=\epsilon_{\rm mis}^{\rm B}\,.
\label{eq:epsmis}
\end{equation}
The mean site energies $\bar\epsilon_\upalpha^\upalpha$ of
$\upalpha$ ions in the region of the $\epsilon_{\rm A}\epsilon_{\rm B}$-plane
 favorable for $\upalpha$ ions are equal and set to zero, $\mAA=\mBB=0$. 
 The Pearson correlation coefficients are chosen to be one half, $\rho_{\rm A}=\rho_{\rm B}=1/2$.
We use $\Delta_0$ as the energy unit, yielding a scaled temperature
\begin{equation}
\theta\equiv\frac{\kb T}{\Delta_0}\,.
\end{equation}
The attempt frequency $\nu_{\rm A}=\nu_{\rm B}\equiv\nu$ gives the conductivity scale
\begin{equation}
\sigma_0\equiv \frac{\nu e^2}{a \Delta_0}\,.
\end{equation}
It corresponds to the conductivity of independent ions with charge $e$
and concentration one per lattice site that jump with rate $\nu$ between nearest neighbor sites at 
temperature $\kb T=\Delta_0$ ($\theta=1)$. 

For the saddle point energies 
$\epsilon_{ij}^{{\rm s},\upalpha}$ in Eq.~\eqref{eq:wij0}, we can write
\begin{equation}
\epsilon_{ij}^{{\rm s},\upalpha}=\max(\epsilon_i^\upalpha,\epsilon_j^\upalpha)+b_{ij}^\upalpha\,,
\label{eq:epsij-bij}
\end{equation}
where $b_{ij}^\upalpha$ is the lower barrier for jumps of $\upalpha$ ions between sites $i$ and $j$.
The distribution of the $b_{ij}^\upalpha$ should have 
support at small $b_{ij}^\upalpha\simeq0$ to account for the fact that isotherms
in measured conductivity spectra tend to merge at high frequencies \cite{Kulkarni/etal:1998}.
We set $b_{ij}^\upalpha=0$ for simplicity, thus avoiding to consider further parameters for the energy landscape description.

The symmetric setting with minimal number of parameters is useful for general investigations: the generic key features of the MAE seen in different ionic transport properties should not depend on a specific choice of parameters or parameter range. For application to measurements, asymmetric sets of parameters need to be considered, where 
specific values are obtained by fitting experimental data \cite{Leiber/etal:2026}. 

We are interested in the generic features here and use the same symmetric setting in the following sections, where we probe the BSEM's capability to capture the salient features of ac conductivities and internal friction of mixed alkali glasses.

\section{Frequency dependence of the MAE}
\begin{figure}[t!]
\centering
\includegraphics[width=0.95\columnwidth]{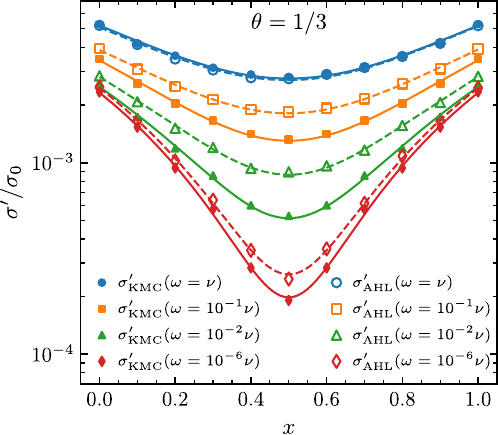}
\caption{\small Real part of the ac ionic conductivity $\sigma'(\omega,T)$ as a function of $x$
at the scaled temperature $\theta=1/3$ and various fixed frequencies $\omega$. 
Parameters are the same as in Fig.~\ref{fig:partial-conductivities}. Full and open symbols mark results from KMC simulations and
the generalized AHL theory, respectively. Solid and dashed lines are drawn as guides to the eye.}
\label{fig:ac-conductivities}
\end{figure}

Figure~\ref{fig:ac-conductivities} shows real parts 
\begin{equation}
\sigma'(\omega,T)=\Re[\hat\sigma_{\rm A}(\omega,T)+\hat\sigma_{\rm B}(\omega,T)]
\end{equation}
of the frequency-dependent total ion conductivity as a function of the mixing ratio $x$ for various frequencies $\omega$ at the scaled temperature $\theta=1/3$. In accordance with experimental findings \cite{Chen/etal:1992, Rathan/etal:2022}, the MAE becomes weaker with increasing frequency. The conductivities $\sigma'(\omega,T)$ calculated with the VAC method
for the dynamics of noninteracting ions in the associated barrier landscapes
yield a very good agreement with the KMC results for high $\omega=\nu$,
corresponding to about 100~GHz in experiments. In fact,
the mapping onto the barrier landscapes yields exact values for 
$\sigma'(\omega,T)$ in the limit $\omega\to\infty$ \cite{Lohmann/etal:2026}.
For the lowest frequency $\omega=10^{-6}\nu$ in Fig.~\ref{fig:ac-conductivities},
the approximation is quite good also, while
for $\omega=10^{-2}\nu$ and $\omega=10^{-3}\nu$ deviations from the KMC results are more pronounced.
The key feature of the MAE weakening with higher frequency is nicely seen also in the approximate calculation. 

The decreasing strength of the MAE with $\omega$ can be understood from the fact 
that the mobile ions explore a space of volume $V(\omega)$ that shrinks with increasing
$\omega$. Accordingly, the site energy fluctuations experienced by the ions become smaller.
For the interacting system, the MAE weakening is particularly transparent for the dynamics of noninteracting ions in the associated barrier landscapes:
the maximum barrier $u_{ij}^\upalpha$
to be surmounted decreases with shrinking volume $V(\omega)$ when the frequency rises -- an effect well known from percolation theory.

Alternatively, the MAE weakening can be understood
from the different types of temperature dependence of the high-frequency conductivity
$\sigma_{\infty}(T)=\sigma'(\omega\to\infty,T)$ and dc conductivity 
$\sigma_{\rm dc}(T)=\sigma'(\omega\to0,T)$ at low $T$: while $\sigma_{\rm dc}(T)$ is thermally activated with the activation energy 
\begin{equation}
E_{\rm dc}=\min(E_{\rm A},E_{\rm B})\,,
\end{equation}
$\sigma_{\infty}(T)$ increases proportional to $T$. Hence, the strong changes of $\sigma_{\rm dc}$ with the mixing ratio over several orders of magnitude at low $T$, caused by the change of $E_{\rm dc}$ with $x$, are not seen in $\sigma_{\infty}(T)$.

The linear increase of $\sigma_{\infty}(T)$ with $T$ follows by writing 
\vspace*{-1ex}\begin{equation}
\sigma_{\infty}(T)
=\frac{e^2}{\kb T a}\,
\left[c_{\rm A}\Gamma_{\rm A}+c_{\rm B}\Gamma_{\rm B}\right]\,,
\label{eq:siginf-Gamma}
\end{equation}
\vspace*{-2ex}where
\begin{align}
\Gamma_\upalpha&=
\int\hspace{-0.3em}\dd\epsilon_1^{\rm A}\hspace{-0.3em}
\int\hspace{-0.3em}\dd\epsilon_1^{\rm B}\hspace{-0.3em}
\int\hspace{-0.3em}\dd\epsilon_2^{\rm A}\hspace{-0.3em}
\int\hspace{-0.3em}\dd\epsilon_2^{\rm B}\,
\psi(\epsilon_1^{\rm A},\epsilon_1^{\rm B})\,\psi(\epsilon_2^{\rm A},\epsilon_2^{\rm B})
\nonumber\\
&\hspace{4em}
{}\times f_\upalpha(\beta\tilde\epsilon_1^{\rm A},\beta\tilde\epsilon_1^{\rm B})\,
v(\beta\tilde\epsilon_1^{\rm A},\beta\tilde\epsilon_1^{\rm B})\,w_{12}^{0,\upalpha}
\label{eq:Gammaalpha}
\end{align}
is the mean jump rate of $\upalpha$ ions
between two neighboring sites in thermal equilibrium.
As the chemical potentials increase quadratically with $T$ at low temperatures,
\begin{equation}
\mu_\upalpha(T,c_{\rm A},c_{\rm B})\sim \epsilon_{\rm F}^\upalpha(c_{\rm A}, c_{\rm B})+
\zeta_\upalpha(\kb T)^2\,,
\end{equation}
with constants $\zeta_\upalpha$ given in \cite{Leiber/etal:2026}, it is straightforward to
show that 
\begin{equation}
\Gamma_\upalpha\sim T^2 
\end{equation}
for $T\to0$, yielding  $\sigma_{\infty}(T)\sim T$ from Eq.~\eqref{eq:siginf-Gamma}.
This behavior does not change if a distribution of barriers $b_{ij}^\upalpha$ with 
support at $b_{ij}\simeq0$ in Eq.~\eqref{eq:epsij-bij}
is taken into account.

\begin{figure}[t!]
\centering
\includegraphics[width=\columnwidth]{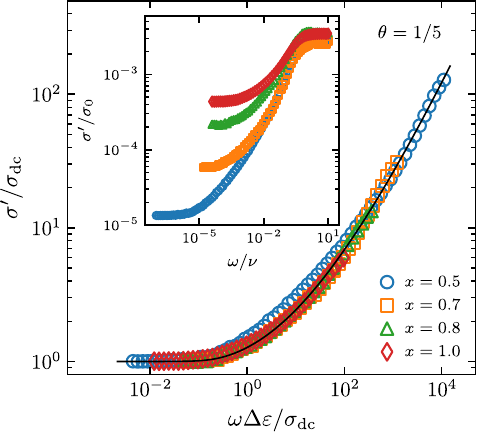}
\caption{\small Scaled conductivity spectra from the BSEM for different mixing ratios $x$ at temperature $\theta=1/5$. 
Parameters are the same as in Fig.~\ref{fig:partial-conductivities}. The inset shows the unscaled spectra. 
Frequencies $\omega\lesssim10^{-1}\nu$ are taken into account in the scaled spectra. 
The solid line marks the scaling function of the random barrier model (see text).}
\label{fig:conductivity_spectra}
\end{figure}

In single ionic glasses, conductivity spectra obey the time-temperature superposition principle, i.e.\ they can be scaled onto a common master curve by normalizing the conductivity $\hat\sigma(\omega,T)$ to the dc conductivity and
by a proper scaling of the frequency \cite{Summerfield:1985, Roling/etal:1997, Sidebottom:1999, Ghosh/Pan:2000, Dyre/Schroeder:2000, Lohmann/etal:2026}. 
Given that
the conductivity $\hat\sigma(\omega,T)$ has a regular low-frequency expansion \cite{Schroeder/Dyre:2000},
it was reasoned that 
the scaled frequency should be $\omega\Delta\varepsilon/\sigma_{\rm dc}$, where $\Delta\varepsilon=\varepsilon'(0)-\varepsilon_\infty$ is the dielectric strength, i.e.\ the difference between the real part of the dielectric function at zero and high frequency where only electronic polarization is present.
When trying to extend this scaling to conductivity spectra at different mixing ratios, it was found
that is not perfect.
The scaled spectra in particular showed a missing collapse onto a common master curve close to the onset of conductivity dispersion.

Figure~\ref{fig:conductivity_spectra} shows scaled conductivity spectra of the BSEM. 
The unscaled spectra are shown in the inset
up to frequencies, where the conductivity reaches the limiting value $\sigma_\infty$ according to Eqs.~\eqref{eq:siginf-Gamma}, \eqref{eq:Gammaalpha}. 
Experimentally, a plateau at high frequencies is generally not seen because at its
onset at $\omega\simeq\nu$ vibrational motions of the ions in their potential minima become relevant, which are not considered in the hopping dynamics. Therefore only frequencies $\omega\lesssim10^{-1}\nu$ are taken into account in the scaled spectra. The solid line marks the scaling function of the random barrier model 
given by the implicit equation 
$\Sigma\ln\Sigma=i\tilde\omega[1+2.66\,i(\tilde\omega/\Sigma)]^{-1/3}$, where 
$\Sigma=\Sigma(\omega,T)=\hat\sigma(\omega,T)/\sigma_{\rm dc}(T)$ is the normalized complex ac conductivity
and $\tilde\omega=\omega\Delta\varepsilon/\sigma_{\rm dc}$ the scaled frequency \cite{Schroeder/Dyre:2008}.

The missing data collapse can be understood from the fact that with varying $x$, the energy landscape changes and accordingly the length scales beyond which the dynamics of the more mobile ion become diffusive at long times.
The respective length scales determine different crossover frequencies to the regimes of dc transport.
Since not only one characteristic frequency is present, 
no perfect scaling behavior is observed.

\section{Internal friction}
The internal friction induced by ionic motion in mixed alkali glasses shows quite
peculiar features. Typically, it is determined by measuring 
the damping of a torsional glass fiber pendulum as a function of temperature at fixed fiber length, corresponding to a fixed eigenfrequency of the pendulum \cite{Shelby:1965, Shelby/Day:1969, Shelby/Day:1970, McVay/Day:1970, Sakamura/etal:2003}. Experimental internal friction spectra at fixed temperatures were reported, for example, 
in Refs.~\cite{Chocinski-Arnault/Riviere:2007, Riviere/Chocinski-Arnault:2009}.
The main findings can be summarized as follows \cite{Zdaniewski/etal:1979}:

\begin{list}{--}{\setlength{\leftmargin}{1.2em}\setlength{\rightmargin}{0em}
\setlength{\itemsep}{0ex}\setlength{\topsep}{0ex}}

\item A single-alkali peak appears that is observable also in the single ionic glasses at $x=0,1$. Its peak frequency
$\omega_{\rm s}$ is activated with an activation energy close to that of the mobility (or partial conductivity) of the more mobile ion.

\item An additional mixed-alkali peak emerges at a peak frequency $\omega_{\rm m}$ that
has an activation energy with no obvious relation to the activation energies of the ion mobilities. 

\item The height of the mixed-alkali peak can be comparable to that of the single-alkali peak already at small substititions $x\lesssim0.1$ or $x\gtrsim0.9$
of the majority ion. It becomes the larger the more similar the two types of ions are, as, for example, quantified by the ion radii.

\item In glasses with $m>2$ different types of mobile ions, the number of mixed peaks can be as large as the number $m(m\!-\!1)/2$ of distinct pairings of the mobile ion types.

\end{list}

Theoretically, the internal friction was studied previously in 
Ref.~\cite{Peibst/etal:2005, Maass/Peibst:2006}. By considering 
the change of the site energies in the presence of a weak oscillating shear field, 
the heat generation per oscillation period was derived using linear response theory. 
The averaged ratio of the dissipated heat to the stored mechanical energy gives the internal friction $Q^{-1}(\omega,T)$.
Under the simplifying assumptions that
the coupling constants of the shear field to the A and B site energies are the same on average and
spatially uncorrelated, a simple expression for 
$Q^{-1}(\omega,T)$ is obtained:
\begin{equation}
Q^{-1}(\omega,T)=\frac{\kappa\omega}{k_{\rm B}T}\Bigl[
\sum_{\upalpha=\rm A,B}S_{\upalpha\rm V}(\omega,T)
+2S_{\rm AB}(\omega,T)\Bigr]\,.
\label{eq:intfr}
\end{equation}
Here, $\kappa$ is a constant with dimension of an energy and 
\begin{equation}
S_{\upalpha\upbeta}(\omega,T)=\int\limits_0^\infty\hspace{-0.2em}
dt\, C_{\upalpha\upbeta}(t,T)\cos(\omega t)
\label{eq:salphabeta}
\end{equation}
are Fourier cosine transforms of the correlations functions
\begin{equation}
C_{\upalpha\upbeta}(t,T)=\frac{1}{M}\sum_{i=1}^M 
\langle n_i^\upalpha(t) n_i^\upbeta(0)\rangle_{\rm eq}\,.
\label{eq:Calphabeta}
\end{equation}
As the jump rates \eqref{eq:wij0} in equilibrium fullfill the condition of detailed balance,
the $C_{\upalpha\upbeta}(t,T)$ are symmetric and invariant under time reversal,
\begin{equation}
C_{\upalpha\upbeta}(t,T)=C_{\upbeta\upalpha}(t,T)=C_{\upalpha\upbeta}(-t,T)\,,
\end{equation}
implying $S_{\upalpha\upbeta}(\omega,T)=S_{\upbeta\upalpha}(\omega,T)$.

\begin{figure}[t!]
\centering
\includegraphics[width=\columnwidth]{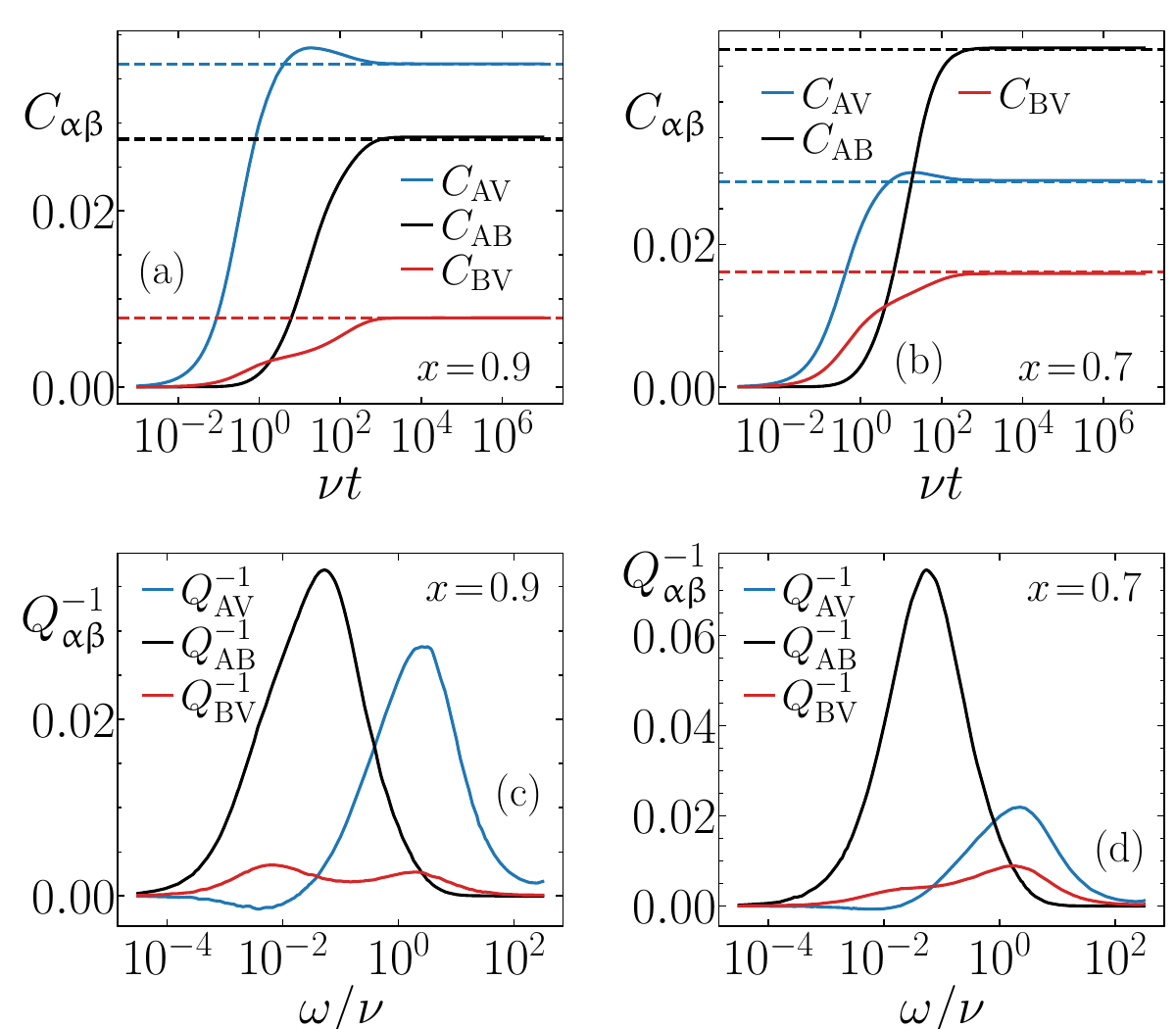}
\caption{\small (a),(b) Simulated correlation functions $C_{\upalpha\upbeta}(t,T)$
quantifying the likelihood of exchanges between $\upalpha$ and $\upbeta$ species 
on a site in time $t$ [Eq.~\eqref{eq:Calphabeta}] for two mixing ratios $x=0.9$ and $x=0.7$
at the scaled temperature $\theta=1$.
The dashed horizontal lines mark the long-time limits $C_{\upalpha\upbeta}^\infty(T)$ 
[Eqs.~\eqref{eq:CinfalphaV}, \eqref{eq:CinfAB}].
(c),(d) Corresponding contributions $Q^{-1}_{\upalpha\upbeta}(\omega,T)$ to the internal friction
calculated from Eqs.~\eqref{eq:intfr}, \eqref{eq:salphabeta}. The vacancy concentration is $c_{\rm V}=0.05$. All other
parameters are the same as in Fig.~\ref{fig:partial-conductivities}.}
\label{fig:correlations_internal_friction}
\end{figure}

The correlation function $C_{\rm AB}(t,T)$ quantifies the likelihood of an exchange between A 
and B ions in time $t$ at the same site in thermal equilibrium. 
This process gives rise to the mixed-alkali internal friction peak.
The correlation functions $C_{\upalpha{\rm V}}(t,T)$ quantify the likelihood of an exchange between $\upalpha$ ions and vacancies V at the same site in time $t$. They lead to single-alkali peaks, where in the total spectrum $Q^{-1}(\omega,T)$ only the peak of the more mobile ion is visible. This is because the
single-alkali peak of the less mobile ion is small and masked by the other single-alkali and the mixed-alkali peak.

Figure~\ref{fig:correlations_internal_friction} shows the correlation functions together with
their contributions
\begin{subequations}
\label{eq:Qcontributions}
\begin{align}
Q_{\upalpha{\rm V}}^{-1}(\omega,T)&=\frac{\kappa\omega}{k_{\rm B}T}S_{\upalpha{\rm V}}(\omega,T)\,,
\label{eq:QalphaV}\\
Q_{\rm AB}^{-1}(\omega,T)&=2\frac{\kappa\omega}{k_{\rm B}T}S_{\rm AB}(\omega,T)
\label{eq:QAB}\,,
\end{align}
\end{subequations}
to the total spectrum in Eq.~\eqref{eq:intfr} for two mixing ratios $x=0.9$ and 0.7.
We have set $\kappa=\Delta_0$.
At long times, the $C_{\upalpha\upbeta}(t,T)$ approach the constants 
$C_{\upalpha\upbeta}^\infty(T)\equiv C_{\upalpha\upbeta}(t\!\to\!\infty,T)=
\sum_{i=1}^M \langle n_i^\upalpha\rangle_{\rm eq}\langle n_i^\upbeta\rangle_{\rm eq}/M$ 
given by
\begin{subequations}
\label{eq:Cinf}
\begin{align}
C_{\upalpha{\rm V}}^\infty(T)
&=\int\hspace{-0.3em}\dd\epsilon_{\rm A}\hspace{-0.3em}
\int\hspace{-0.3em}\dd\epsilon_{\rm B}\,
\psi(\epsilon_{\rm A},\epsilon_{\rm B})\,\label{eq:CinfalphaV}\\
&\hspace{6.5em}\times{}f_\upalpha(\beta\tilde\epsilon_{\rm A},\beta\tilde\epsilon_{\rm B})
v(\beta\tilde\epsilon_{\rm A},\beta\tilde\epsilon_{\rm B})\,,\nonumber\\
C_{\rm AB}^\infty(T)
&=\int\hspace{-0.3em}\dd\epsilon_{\rm A}\hspace{-0.3em}
\int\hspace{-0.3em}\dd\epsilon_{\rm B}\,
\psi(\epsilon_{\rm A},\epsilon_{\rm B})\,\label{eq:CinfAB}\\
&\hspace{6em}\times{}f_{\rm A}(\beta\tilde\epsilon_{\rm A},\beta\tilde\epsilon_{\rm B})
f_{\rm B}(\beta\tilde\epsilon_{\rm A},\beta\tilde\epsilon_{\rm B})\,.\nonumber
\end{align}
\end{subequations}
These constants give rise to a $\delta(\omega)$-contribution in the Fourier transforms
$S_{\upalpha\upbeta}(\omega,T)$, which
are irrelevant for the $Q_{\upalpha\upbeta}^{-1}(\omega,T)$ in
Eqs.~\eqref{eq:QalphaV}, \eqref{eq:QAB} due to the prefactor $\propto\omega$. 
One may
replace the $n_i^\upalpha(t)$ by 
$\delta n_i^\upalpha(t)=n_i^\upalpha(t)-\langle n_i^\upalpha\rangle_{\rm eq}$
in Eq.~\eqref{eq:Calphabeta} if one prefers to deal with correlation functions decaying to zero in time.

\begin{table}[b]
\caption{\label{tab:taualphabeta}Average times in units of $\nu^{-1}$ for the exchange between A ions and vacancies V, and between  A and B ions
on the same site in thermal equilibrium [Eq.~\eqref{eq:taualphabeta}]. Parameters are the same as in Fig.~\ref{fig:correlations_internal_friction}.}
\begin{ruledtabular}
\begin{tabular}{l c c c c c c c c c c c c}
 $x$ & 1 & 0.9 & 0.8 & 0.7 & 0.6 & 0.5 \\
\hline
$\tau_{\rm AV}$ & 0.27 & 0.29 & 0.31 & 0.35 & 0.40 & 0.48 \\
$\tau_{\rm AB}$ & -- & 17.14 & 14.03 & 12.29 & 11.58 & 11.28 \\
\end{tabular}
\end{ruledtabular}
\end{table}

The approach of the $C_{\upalpha\upbeta}(t,T)$ towards the $C_{\upalpha\upbeta}^\infty(T)$
is fastest for the A-V
process, i.e.\ the exchange of the more mobile ion with vacancies. The A-B
exchange is slower on average. 
Characteristic times $\tau_{\upalpha\upbeta}$ for the exchange processes can be defined by
requiring the correlation functions to reach half of their long-time limit, i.e.\ by
\begin{equation}
C_{\upalpha\upbeta}(\tau_{\upalpha\upbeta},T)=\frac{1}{2}\,C_{\upalpha\upbeta}^\infty(T)\,.
\label{eq:taualphabeta}
\end{equation}
The times $\tau_{\rm AV}$ and $\tau_{\rm AB}$
are listed in Table~\ref{tab:taualphabeta} and
satisfy $\tau_{\rm AV}<\tau_{\rm AB}$.
Accordingly, the single-alkali contribution $Q_{\rm AV}^{-1}(\omega,T)$ has a peak at a frequency
$\omega_{\rm s}\simeq 1/\tau_{\rm AV}$
higher than the peak frequency $\omega_{\rm m}\simeq 1/\tau_{\rm AB}$
of the mixed-alkali contribution $Q_{\rm AB}^{-1}(\omega,T)$, see Figs.~\ref{fig:correlations_internal_friction}(b),(d). The height of the mixed-alkali
peak at $x=0.9$ is already comparable to that of the single-alkali peak, and at 
$x=0.7$ it is more than three times larger.

The single-alkali contribution 
$Q_{\rm BV}^{-1}(\omega,T)$ of the B ions shows two shallow maxima, suggesting that actually two characteristic times are involved in the exchange between the less mobile ions and vacancies. A two-fold process can be seen also in $C_{\rm BV}(t,T)$ in
Figs.~\ref{fig:correlations_internal_friction}(a),(c).
However, $Q_{\rm BV}^{-1}(\omega,T)$ is much smaller that the two other contributions. It
would not be resolvable from the total spectrum $Q^{-1}(\omega,T)$ accessible in experiments.

\begin{figure}[t!]
\centering
\includegraphics[width=\columnwidth]{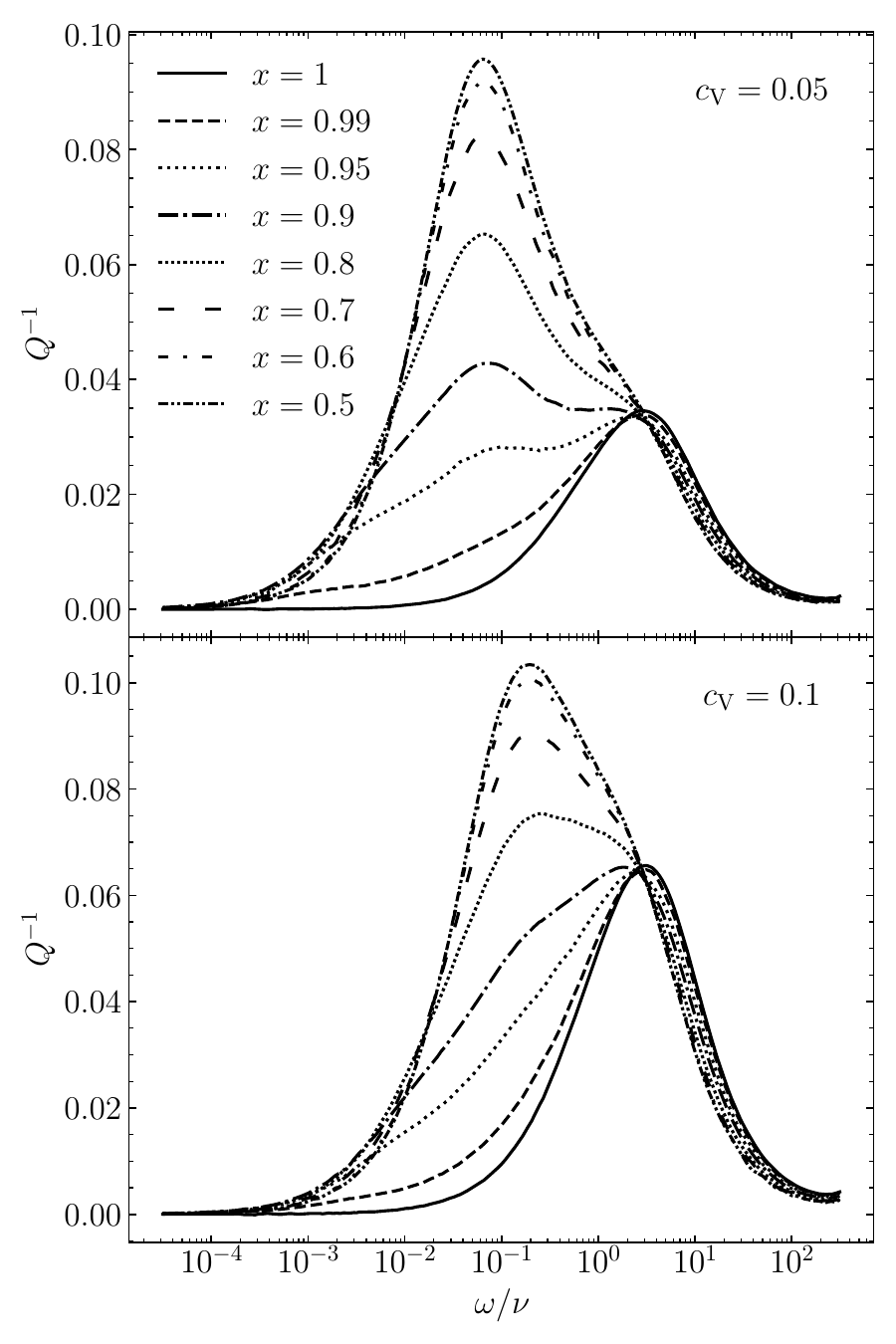}
\caption{\small Internal friction spectra for (a) $c_{\rm V}=0.05$ and (b) $c_{\rm V}=0.1$ and various mixing ratios $x$ at the scaled temperature $\theta=1$.
Parameters are the same as in Fig.~\ref{fig:partial-conductivities}.}
\label{fig:sum_spectrum}
\end{figure}

Figure~\ref{fig:sum_spectrum}(a) shows the total spectrum $Q^{-1}(\omega,T)$ for various mixing ratios $x$. At $x=1$, only the single-alkali peak appears at the frequency $\omega_{\rm s}\simeq2\nu$ with symmetric shape in the log-linear representation.
Already at $x=0.99$, i.e.\ 1\% substitution of A by B ions,
it has a pronounced asymmetry due to
an extended wing left from the maximum, which
indicates the presence of the mixed-alkali contribution. This contribution develops into a peak at the frequency $\omega_{\rm m}\simeq 10^{-1}\nu$ 
for $x\gtrsim0.95$. With decreasing $x$ in the range $1\ge x\ge 0.5$, the
mixed-alkali peak becomes higher and the single-alkali peak lower.
For $x\gtrsim0.8$, the mixed-alkali peak dominates the spectrum and the single-alkali contribution appears as a shoulder at high frequencies. This form of the spectrum is similar to the one observed for mixed alkali
silicate glasses $(1\!-\!x)$Na$_2$O-$x$Li$_2$O-3SiO$_2$ in 
Refs.~\cite{Chocinski-Arnault/Riviere:2007, Riviere/Chocinski-Arnault:2009}.
The frequencies considered here are, however, much larger than in the experiment when
assuming the attempt frequency to be of the order $10^{12}\,\si{Hz}$, corresponding to
typical phonon frequencies.

\begin{figure}[t!]
\centering
\includegraphics[width=0.8\columnwidth]{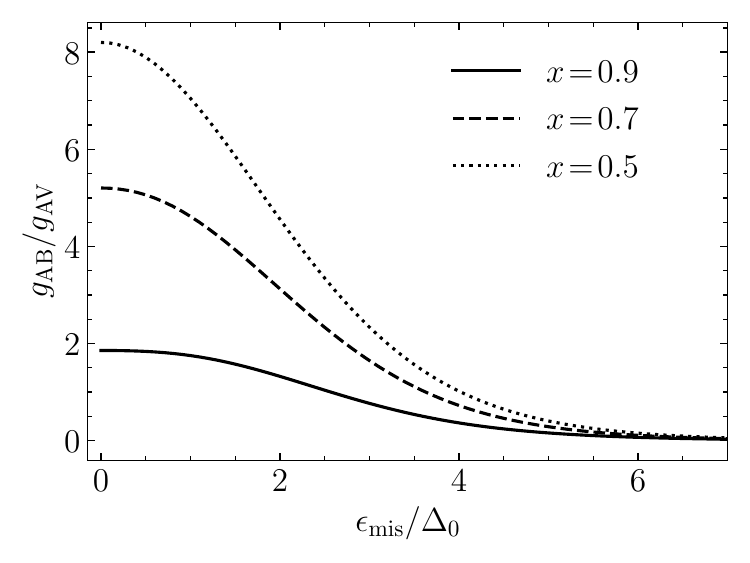}
\caption{\small Fraction $g_{\rm AB}/g_{\rm AV}$ as a function
of the mismatch energy $\epsilon_{\rm mis}$ (in units of $\Delta_0)$ for three mixing ratios $x$
at the scaled temperature $\theta=1$ and vacancy concentration $c_{\rm V}=0.05$. 
Parameters are the same as in Fig.~\ref{fig:partial-conductivities}.
The coefficients $g_{\upalpha\upbeta}=
\sum_{i=1}^M \langle n_i^\upalpha\rangle_{\rm eq}\langle n_i^\upbeta\rangle_{\rm eq}/(Mc_\upalpha c_\upbeta)
=C_{\upalpha\upbeta}^\infty(T)/(c_\upalpha c_\upbeta)$
quantify the correlation between site occupation probabilities [cf.\ Eq.~\eqref{eq:C-g}].
The data were calculated by numerical evaluation of the integrals in Eqs.~\eqref{eq:CinfalphaV} and \eqref{eq:CinfAB}.}
\label{fig:ratio_fermi_factor}
\end{figure}

The possibility to observe very pronounced mixed-alkali peaks already at small substitutions of the majority ion by the minority ion is a remarkable feature. It requires the concentration $c_{\rm V}$ of vacancies to be low and the mismatch between the ion types, as quantified by the mismatch energy $\epsilon_{\rm mis}$ in Eq.~\eqref{eq:epsmis}, to be small.
To show this, let us introduce the normalized correlation functions $\tilde C_{\upalpha\upbeta}(t,T)\!=\!
C_{\upalpha\upbeta}(t,T)/C_{\upalpha\upbeta}^\infty(T)$ and their
Fourier cosine transforms
$\tilde S_{\upalpha\upbeta}(\omega,T)=S_{\upalpha\upbeta}(\omega,T)/C_{\upalpha\upbeta}^\infty(T)$.
The respective contributions $\tilde Q^{-1}_{\upalpha{\rm V}}\!=\!(\kappa \omega/\kb T) \tilde S_{\upalpha{\rm V}}$ and $\tilde Q^{-1}_{\rm AB}\!=\!2(\kappa \omega/\kb T) \tilde S_{\rm AB}$ to the internal friction then have comparable intensity.
Setting
\begin{equation}
C_{\upalpha\upbeta}^\infty(T)=\frac{1}{M}\sum_{i=1}^M \langle n_i^\upalpha\rangle_{\rm eq}\langle n_i^\upbeta\rangle_{\rm eq}=c_\upalpha c_\upbeta g_{\upalpha\upbeta}(T)\,,
\label{eq:C-g}
\end{equation}
where $g_{\upalpha\upbeta}(T)=C_{\upalpha\upbeta}^\infty(T)/C_{\upalpha\upbeta}^\infty(T\!\to\!\infty)$
with  $C_{\upalpha\upbeta}^\infty(T\to\infty)\!=\!c_\upalpha c_\upbeta$
is a coefficient quantifying the correlation between site occupation probabilities, we
can write
\begin{subequations}
\label{eq:Qcontributions}
\begin{align}
Q_{\upalpha{\rm V}}^{-1}(\omega,T)&=c_\upalpha c_{\rm V}g_{\upalpha{\rm V}}(T)
\tilde Q^{-1}_{\upalpha{\rm V}}(\omega,T)\,,
\label{eq:QalphaV}\\
Q_{\rm AB}^{-1}(\omega,T)&=c_{\rm A}c_{\rm B}g_{\rm AB}(T)\tilde Q^{-1}_{\rm AB}(\omega,T)
\label{eq:QAB}\,.
\end{align}
\end{subequations}
Since peaks of the $\tilde Q^{-1}_{\upalpha\upbeta}(\omega,T)$ have comparable heights,
the difference in heights of the single-alkali and mixed-alkali peak is mainly given by
the prefactors $c_{\rm A}c_{\rm V}g_{\rm AV}(T)$ and $c_{\rm A}c_{\rm B}g_{\rm AB}(T)$
(if A is the more mobile ion). Accordingly, for the mixed-alkali peak to be comparable to the single-alkali one for small $c_{\rm B}$, $c_{\rm V}$ has to be small also. 

Differently speaking, for larger $c_{\rm V}$ the single-alkali peak is more prevalent.
This is demonstrated in Fig.~\ref{fig:sum_spectrum}(b), where the total spectrum $Q^{-1}
(\omega,T)$ is shown for the same parameters as in Fig.~\ref{fig:sum_spectrum}(a) but a two 
times larger vacancy concentration $c_{\rm V}=0.1$. 

As $C_{\rm AB}^\infty(T)$ rises if the disorder-averaged product of 
probabilities of occupying a site by an A and B ion increases, the mixed-alkali peak becomes the larger the more similar
the two types of mobile ions are, i.e.\ the smaller $\epsilon_{\rm mis}$. 
Specifically, when considering $c_{\rm B}=c_{\rm V}$ in Eqs.~\eqref{eq:QalphaV}, 
\eqref{eq:QAB}, the fraction $g_{\rm AB}/g_{\rm AV}$ gives an estimate of the strength
of the mixed-alkali relative to the single-alkali peak. This fraction is readily
calculated by numerical evaluation of the integrals in Eqs.~\eqref{eq:CinfalphaV} and
\eqref{eq:CinfAB}.
Results for $g_{\rm AB}/g_{\rm AV}$ as function of $\epsilon_{\rm mis}$ are shown in 
Fig.~\ref{fig:ratio_fermi_factor} and demonstrate that pronounced mixed-alkali peaks occur for small $\epsilon_{\rm mis}$.

Moreover, when generalizing the theory to $m>2$ types of mobile ions, there are 
$m(m\!-\!1)/2$ possibilities of exchanging one type of mobile ion with another type at a site.
Thus the number of mixed-alkali peaks can be as large as the number of distinct pairings of the mobile ion types.

\section{Conclusions}
The recently introduced BSEM provides a theoretical description of the MAE consistent with the thermodynamics and kinetics of mixtures. We here showed that the BSEM is not only capable to capture the salient features of dc transport properties but also of ac conductivities and the internal friction. 

We have calculated ac ionic conductivities of the BSEM by both KMC simulations and with the generalized AHL theory, where the dynamics of interacting ions are mapped onto that of noninteracting ions in barrier landscapes, one for each ion type. Using the AHL theory, approximate results for partial conductivities and their activation energies 
are obtained in very short computing times of the order of minutes on a standard personal computer.

The results show a weakening of the MAE with increasing frequency in accordance with measurements. 
At very high frequencies, the ac conductivity is no longer thermally activated.
Conductivity spectra for different mixing ratios cannot be successfully 
scaled. As in experiments, the data do not fall onto a common curve close to the onset of conductivity dispersion. 
The BSEM allows us to understand the deviations from scaling: the energy landscape changes with $x$ and therefore the characteristic length scale on which the dynamics of the
more mobile ions becomes memoryless, i.e.\ where it reaches its long-time diffusion coefficient.
As a consequence, there is not only one characteristic length scale that determines a crossover frequency between the dispersive and the dc regime. 

We moreover conjecture that the time-temperature scaling of conductivity spectra can become worse
compared to that of single ionic glasses at mixing ratios, where the partial conductivities are comparable. This is suggested by the mapping to the distinct barrier landscapes: here a single characteristic
frequency can be missing because 
the length scales for reaching long-time diffusion are different for the two types of ions.

Internal friction spectra predicted by the BSEM were calculated based on a theory developed earlier \cite{Maass/Peibst:2006}. They are in qualitative agreement with experimental observations. The processes responsible for the emergence of internal friction peaks are changes in the occupation states of ion sites. The single-alkali peak is caused by the process where the state of a site changes between vacant and occupied by the more mobile ion.
Responsible for the mixed-alkali peak is the process, where sites originally occupied by one type of mobile ion A are
occupied by the other type B at a later time. This A-B exchange process is slower than the A-V exchange process, causing the mixed-alkali peaks to occur at lower frequencies than the single-alkali peak. 
We showed that a small concentration of vacancies allows
the mixed-alkali peak to become large already for small substitutions of the majority by the minority ion, i.e.\ at mixing ratios close to zero or one. For more similar ion types,
A-B-exchange processes are more frequent, causing mixed-alkali internal friction peaks to become more
pronounced. For glasses containing more than two types of mobile ions, exchange processes between all distinct pairs of ion types are possible. Accordingly, a corresponding number of possible mixed-alkali internal friction peaks can occur.

We have presented first results here for ac conductivities and internal friction predicted by
the BSEM, focusing on the generic features found in experiments. In the future, we aim to develop the theory further
to allow for an efficient fitting of experimental data, as we
have demonstrated already for dc partial conductivities in Ref.~\cite{Leiber/etal:2026}. For ac conductivities, a suitable approach is already available based on the generalized AHL theory and the VAC method for calculating ac conductivities 
in the barrier landscape description. For the internal friction, an efficient numerical method needs to be designed for determining site exchange correlation functions in the barrier landscapes.

\begin{acknowledgments}
We gratefully acknowledge financial support by the
Deutsche Forschungsgemeinschaft (Project No.\ 428906592), and the use
of a high-performance computing cluster funded by the Deutsche
Forschungsgemeinschaft (Project No.\ 456666331).
\end{acknowledgments}


\begin{thebibliography}{60}%
\makeatletter
\providecommand \@ifxundefined [1]{%
 \@ifx{#1\undefined}
}%
\providecommand \@ifnum [1]{%
 \ifnum #1\expandafter \@firstoftwo
 \else \expandafter \@secondoftwo
 \fi
}%
\providecommand \@ifx [1]{%
 \ifx #1\expandafter \@firstoftwo
 \else \expandafter \@secondoftwo
 \fi
}%
\providecommand \natexlab [1]{#1}%
\providecommand \enquote  [1]{``#1''}%
\providecommand \bibnamefont  [1]{#1}%
\providecommand \bibfnamefont [1]{#1}%
\providecommand \citenamefont [1]{#1}%
\providecommand \href@noop [0]{\@secondoftwo}%
\providecommand \href [0]{\begingroup \@sanitize@url \@href}%
\providecommand \@href[1]{\@@startlink{#1}\@@href}%
\providecommand \@@href[1]{\endgroup#1\@@endlink}%
\providecommand \@sanitize@url [0]{\catcode `\\12\catcode `\$12\catcode
  `\&12\catcode `\#12\catcode `\^12\catcode `\_12\catcode `\%12\relax}%
\providecommand \@@startlink[1]{}%
\providecommand \@@endlink[0]{}%
\providecommand \url  [0]{\begingroup\@sanitize@url \@url }%
\providecommand \@url [1]{\endgroup\@href {#1}{\urlprefix }}%
\providecommand \urlprefix  [0]{URL }%
\providecommand \Eprint [0]{\href }%
\providecommand \doibase [0]{https://doi.org/}%
\providecommand \selectlanguage [0]{\@gobble}%
\providecommand \bibinfo  [0]{\@secondoftwo}%
\providecommand \bibfield  [0]{\@secondoftwo}%
\providecommand \translation [1]{[#1]}%
\providecommand \BibitemOpen [0]{}%
\providecommand \bibitemStop [0]{}%
\providecommand \bibitemNoStop [0]{.\EOS\space}%
\providecommand \EOS [0]{\spacefactor3000\relax}%
\providecommand \BibitemShut  [1]{\csname bibitem#1\endcsname}%
\let\auto@bib@innerbib\@empty
\bibitem [{\citenamefont {Isard}(1969)}]{Isard:1969}%
  \BibitemOpen
  \bibfield  {author} {\bibinfo {author} {\bibfnamefont {J.~O.}\ \bibnamefont
  {Isard}},\ }\bibfield  {title} {\bibinfo {title} {The mixed alkali effect in
  glass},\ }\href
  {https://doi.org/https://doi.org/10.1016/0022-3093(69)90003-9} {\bibfield
  {journal} {\bibinfo  {journal} {J. Non-Cryst. Solids}\ }\textbf {\bibinfo
  {volume} {1}},\ \bibinfo {pages} {235 } (\bibinfo {year} {1969})}\BibitemShut
  {NoStop}%
\bibitem [{\citenamefont {Day}(1976)}]{Day:1976}%
  \BibitemOpen
  \bibfield  {author} {\bibinfo {author} {\bibfnamefont {D.~E.}\ \bibnamefont
  {Day}},\ }\bibfield  {title} {\bibinfo {title} {Mixed alkali glasses -- their
  properties and uses},\ }\href
  {https://doi.org/https://doi.org/10.1016/0022-3093(76)90026-0} {\bibfield
  {journal} {\bibinfo  {journal} {J. Non-Cryst. Solids}\ }\textbf {\bibinfo
  {volume} {21}},\ \bibinfo {pages} {343 } (\bibinfo {year}
  {1976})}\BibitemShut {NoStop}%
\bibitem [{\citenamefont {Ingram}(1987)}]{Ingram:1987}%
  \BibitemOpen
  \bibfield  {author} {\bibinfo {author} {\bibfnamefont {M.~D.}\ \bibnamefont
  {Ingram}},\ }\bibfield  {title} {\bibinfo {title} {Ionic conductivity in
  glass},\ }\href@noop {} {\bibfield  {journal} {\bibinfo  {journal} {Phys.
  Chem. Glasses}\ }\textbf {\bibinfo {volume} {28}},\ \bibinfo {pages} {215}
  (\bibinfo {year} {1987})}\BibitemShut {NoStop}%
\bibitem [{\citenamefont {Dyre}\ \emph {et~al.}(2009)\citenamefont {Dyre},
  \citenamefont {Maass}, \citenamefont {Roling},\ and\ \citenamefont
  {Sidebottom}}]{Dyre/etal:2009}%
  \BibitemOpen
  \bibfield  {author} {\bibinfo {author} {\bibfnamefont {J.~C.}\ \bibnamefont
  {Dyre}}, \bibinfo {author} {\bibfnamefont {P.}~\bibnamefont {Maass}},
  \bibinfo {author} {\bibfnamefont {B.}~\bibnamefont {Roling}},\ and\ \bibinfo
  {author} {\bibfnamefont {D.~L.}\ \bibnamefont {Sidebottom}},\ }\bibfield
  {title} {\bibinfo {title} {Fundamental questions relating to ion conduction
  in disordered solids},\ }\href
  {https://doi.org/{10.1088/0034-4885/72/4/046501}} {\bibfield  {journal}
  {\bibinfo  {journal} {Rep. Prog. Phys.}\ }\textbf {\bibinfo {volume}
  {{72}}},\ \bibinfo {pages} {046501} (\bibinfo {year} {{2009}})}\BibitemShut
  {NoStop}%
\bibitem [{\citenamefont {Bartholomew}(1973)}]{Bartholomew:1973}%
  \BibitemOpen
  \bibfield  {author} {\bibinfo {author} {\bibfnamefont {R.~F.}\ \bibnamefont
  {Bartholomew}},\ }\bibfield  {title} {\bibinfo {title} {Electrical properties
  of phosphate glasses},\ }\href
  {https://doi.org/https://doi.org/10.1016/0022-3093(73)90004-5} {\bibfield
  {journal} {\bibinfo  {journal} {J. Non-Cryst. Solids}\ }\textbf {\bibinfo
  {volume} {12}},\ \bibinfo {pages} {321} (\bibinfo {year} {1973})}\BibitemShut
  {NoStop}%
\bibitem [{\citenamefont {Carette}\ \emph {et~al.}(1983)\citenamefont
  {Carette}, \citenamefont {Ribes},\ and\ \citenamefont
  {Souquet}}]{Carette/etal:1983}%
  \BibitemOpen
  \bibfield  {author} {\bibinfo {author} {\bibfnamefont {B.}~\bibnamefont
  {Carette}}, \bibinfo {author} {\bibfnamefont {M.}~\bibnamefont {Ribes}},\
  and\ \bibinfo {author} {\bibfnamefont {J.}~\bibnamefont {Souquet}},\
  }\bibfield  {title} {\bibinfo {title} {The effects of mixed anions in ionic
  conductive glasses},\ }\href {https://doi.org/10.1016/0167-2738(83)90323-5}
  {\bibfield  {journal} {\bibinfo  {journal} {Solid State Ionics}\ }\textbf
  {\bibinfo {volume} {9--10}},\ \bibinfo {pages} {735} (\bibinfo {year}
  {1983})}\BibitemShut {NoStop}%
\bibitem [{\citenamefont {Machida}\ \emph {et~al.}(1995)\citenamefont
  {Machida}, \citenamefont {Kawachi}, \citenamefont {Ueda}, \citenamefont
  {Shigematsu}, \citenamefont {Nakanishi}, \citenamefont {Takahashi},\ and\
  \citenamefont {Minami}}]{Machida/etal:1995}%
  \BibitemOpen
  \bibfield  {author} {\bibinfo {author} {\bibfnamefont {N.}~\bibnamefont
  {Machida}}, \bibinfo {author} {\bibfnamefont {M.}~\bibnamefont {Kawachi}},
  \bibinfo {author} {\bibfnamefont {A.}~\bibnamefont {Ueda}}, \bibinfo {author}
  {\bibfnamefont {T.}~\bibnamefont {Shigematsu}}, \bibinfo {author}
  {\bibfnamefont {N.}~\bibnamefont {Nakanishi}}, \bibinfo {author}
  {\bibfnamefont {M.}~\bibnamefont {Takahashi}},\ and\ \bibinfo {author}
  {\bibfnamefont {T.}~\bibnamefont {Minami}},\ }\bibfield  {title} {\bibinfo
  {title} {Mixed anion effect of silver ion conducting glasses in the systems
  agi-ag$_2$moo$_4$-ag$_3$po$_4$ and agi-ag$_2$moo$_4$-ag$_2$po$_{3.5}$ and
  structural study by 31{P} {MAS-NMR}},\ }\href
  {https://doi.org/https://doi.org/10.1016/0167-2738(95)00073-F} {\bibfield
  {journal} {\bibinfo  {journal} {Solid State Ionics}\ }\textbf {\bibinfo
  {volume} {79}},\ \bibinfo {pages} {273} (\bibinfo {year} {1995})}\BibitemShut
  {NoStop}%
\bibitem [{\citenamefont {Rau}\ \emph {et~al.}(2001)\citenamefont {Rau},
  \citenamefont {Armand}, \citenamefont {Pradel}, \citenamefont {Varsamis},
  \citenamefont {Kamitsos}, \citenamefont {Granier}, \citenamefont {Ibanez},\
  and\ \citenamefont {Philippot}}]{Rau/etal:2001}%
  \BibitemOpen
  \bibfield  {author} {\bibinfo {author} {\bibfnamefont {C.}~\bibnamefont
  {Rau}}, \bibinfo {author} {\bibfnamefont {P.}~\bibnamefont {Armand}},
  \bibinfo {author} {\bibfnamefont {A.}~\bibnamefont {Pradel}}, \bibinfo
  {author} {\bibfnamefont {C.~P.~E.}\ \bibnamefont {Varsamis}}, \bibinfo
  {author} {\bibfnamefont {E.~I.}\ \bibnamefont {Kamitsos}}, \bibinfo {author}
  {\bibfnamefont {D.}~\bibnamefont {Granier}}, \bibinfo {author} {\bibfnamefont
  {A.}~\bibnamefont {Ibanez}},\ and\ \bibinfo {author} {\bibfnamefont
  {E.}~\bibnamefont {Philippot}},\ }\bibfield  {title} {\bibinfo {title} {Mixed
  cation effect in chalcogenide glasses {Rb$_2$S-Ag$_2$S-GeS$_2$}},\ }\href
  {https://doi.org/10.1103/physrevb.63.184204} {\bibfield  {journal} {\bibinfo
  {journal} {Phys. Rev. B}\ }\textbf {\bibinfo {volume} {63}},\ \bibinfo
  {pages} {184204} (\bibinfo {year} {2001})}\BibitemShut {NoStop}%
\bibitem [{\citenamefont {Bokova}\ \emph {et~al.}(2015)\citenamefont {Bokova},
  \citenamefont {Alekseev},\ and\ \citenamefont {Bychkov}}]{Bokova/etal:2015}%
  \BibitemOpen
  \bibfield  {author} {\bibinfo {author} {\bibfnamefont {M.}~\bibnamefont
  {Bokova}}, \bibinfo {author} {\bibfnamefont {I.}~\bibnamefont {Alekseev}},\
  and\ \bibinfo {author} {\bibfnamefont {E.}~\bibnamefont {Bychkov}},\
  }\bibfield  {title} {\bibinfo {title} {Mixed cation effect in
  {Ag$_2$S-Tl$_2$S-GeS-GeS$_2$} glasses: {C}onductivity and tracer diffusion
  studies},\ }\href {https://doi.org/https://doi.org/10.1016/j.ssi.2014.10.032}
  {\bibfield  {journal} {\bibinfo  {journal} {Solid State Ionics}\ }\textbf
  {\bibinfo {volume} {273}},\ \bibinfo {pages} {55} (\bibinfo {year} {2015})},\
  \bibinfo {note} {the E-MRS Conference, Symposium C "Solid State Ionics: Thin
  Films for Energy and Information Applications"}\BibitemShut {NoStop}%
\bibitem [{\citenamefont {Bokova}\ \emph {et~al.}(2020)\citenamefont {Bokova},
  \citenamefont {Paraskiva}, \citenamefont {Kassem},\ and\ \citenamefont
  {Bychkov}}]{Bokova/etal:2020}%
  \BibitemOpen
  \bibfield  {author} {\bibinfo {author} {\bibfnamefont {M.}~\bibnamefont
  {Bokova}}, \bibinfo {author} {\bibfnamefont {A.}~\bibnamefont {Paraskiva}},
  \bibinfo {author} {\bibfnamefont {M.}~\bibnamefont {Kassem}},\ and\ \bibinfo
  {author} {\bibfnamefont {E.}~\bibnamefont {Bychkov}},\ }\bibfield  {title}
  {\bibinfo {title} {Mixed cation {Ag$_2$S-Tl$_2$S-GeS$_2$} glasses:
  macroscopic properties and {R}aman scattering studies},\ }\href
  {https://doi.org/10.1088/1361-648x/ab7c91} {\bibfield  {journal} {\bibinfo
  {journal} {J. Phys. Condens. Matter}\ }\textbf {\bibinfo {volume} {32}},\
  \bibinfo {pages} {264004} (\bibinfo {year} {2020})}\BibitemShut {NoStop}%
\bibitem [{\citenamefont {Molinelli}\ \emph {et~al.}(1985)\citenamefont
  {Molinelli}, \citenamefont {Tomozawa},\ and\ \citenamefont
  {Takata}}]{Molinelli/etal:1985}%
  \BibitemOpen
  \bibfield  {author} {\bibinfo {author} {\bibfnamefont {J.}~\bibnamefont
  {Molinelli}}, \bibinfo {author} {\bibfnamefont {M.}~\bibnamefont
  {Tomozawa}},\ and\ \bibinfo {author} {\bibfnamefont {M.}~\bibnamefont
  {Takata}},\ }\bibfield  {title} {\bibinfo {title} {Sodium transport in the
  {Na$_2$O-H$_2$O-SiO$_2$} glass system},\ }\href
  {https://doi.org/https://doi.org/10.1111/j.1151-2916.1985.tb09658.x}
  {\bibfield  {journal} {\bibinfo  {journal} {J. Am. Ceram. Soc.}\ }\textbf
  {\bibinfo {volume} {68}},\ \bibinfo {pages} {165} (\bibinfo {year}
  {1985})}\BibitemShut {NoStop}%
\bibitem [{\citenamefont {Garc{\'\i}a}\ \emph {et~al.}(2002)\citenamefont
  {Garc{\'\i}a}, \citenamefont {Ingram},\ and\ \citenamefont
  {Baz{\'a}n}}]{Garcia/etal:2002}%
  \BibitemOpen
  \bibfield  {author} {\bibinfo {author} {\bibfnamefont {N.~J.}\ \bibnamefont
  {Garc{\'\i}a}}, \bibinfo {author} {\bibfnamefont {M.~D.}\ \bibnamefont
  {Ingram}},\ and\ \bibinfo {author} {\bibfnamefont {J.~C.}\ \bibnamefont
  {Baz{\'a}n}},\ }\bibfield  {title} {\bibinfo {title} {Ion transport in
  hydrated sodium silicates (water glasses) of varying water content},\ }\href
  {https://doi.org/10.1016/S0167-2738(01)00990-0} {\bibfield  {journal}
  {\bibinfo  {journal} {Solid State Ionics}\ }\textbf {\bibinfo {volume}
  {146}},\ \bibinfo {pages} {113} (\bibinfo {year} {2002})}\BibitemShut
  {NoStop}%
\bibitem [{\citenamefont {Almeida}\ and\ \citenamefont
  {MacKenzie}(1982)}]{Almeida/Mackenzie:1982}%
  \BibitemOpen
  \bibfield  {author} {\bibinfo {author} {\bibfnamefont {R.~M.}\ \bibnamefont
  {Almeida}}\ and\ \bibinfo {author} {\bibfnamefont {J.~D.}\ \bibnamefont
  {MacKenzie}},\ }\bibfield  {title} {\bibinfo {title} {The electrical
  conductivity of fluorozirconate and chloro-fluorozirconate glasses},\ }\href
  {https://doi.org/10.1007/BF00543884} {\bibfield  {journal} {\bibinfo
  {journal} {J. Mater. Sci.}\ }\textbf {\bibinfo {volume} {17}},\ \bibinfo
  {pages} {2533} (\bibinfo {year} {1982})}\BibitemShut {NoStop}%
\bibitem [{\citenamefont {Kahnt}\ and\ \citenamefont
  {Reau}(1990)}]{Kahnt/Reau:1990}%
  \BibitemOpen
  \bibfield  {author} {\bibinfo {author} {\bibfnamefont {H.}~\bibnamefont
  {Kahnt}}\ and\ \bibinfo {author} {\bibfnamefont {J.~M.}\ \bibnamefont
  {Reau}},\ }\bibfield  {title} {\bibinfo {title} {Mixed mobile ion effect of
  anionic conduction in lead halogenosilicate glasses},\ }\href
  {https://doi.org/10.1002/pssa.2211210109} {\bibfield  {journal} {\bibinfo
  {journal} {Phys. Status Solidi A}\ }\textbf {\bibinfo {volume} {121}},\
  \bibinfo {pages} {73} (\bibinfo {year} {1990})}\BibitemShut {NoStop}%
\bibitem [{\citenamefont {Alenezi}\ \emph {et~al.}(2026)\citenamefont
  {Alenezi}, \citenamefont {Kafle}, \citenamefont {Alsubaie}, \citenamefont
  {Albalawi}, \citenamefont {Pegg},\ and\ \citenamefont
  {Dutta}}]{Alenezi/etal:2026}%
  \BibitemOpen
  \bibfield  {author} {\bibinfo {author} {\bibfnamefont {M.}~\bibnamefont
  {Alenezi}}, \bibinfo {author} {\bibfnamefont {A.~P.}\ \bibnamefont {Kafle}},
  \bibinfo {author} {\bibfnamefont {M.}~\bibnamefont {Alsubaie}}, \bibinfo
  {author} {\bibfnamefont {N.}~\bibnamefont {Albalawi}}, \bibinfo {author}
  {\bibfnamefont {I.~L.}\ \bibnamefont {Pegg}},\ and\ \bibinfo {author}
  {\bibfnamefont {B.}~\bibnamefont {Dutta}},\ }\bibfield  {title} {\bibinfo
  {title} {The mixed halogen-ion effect in lead silicate glasses: A correlative
  study of ionic transport and optical spectroscopy in the
  {45PbO-xPbF$_2$-(20-x)PbCl$_2$-35SiO$_2$} system},\ }\href
  {https://doi.org/10.3390/electronicmat7010003} {\bibfield  {journal}
  {\bibinfo  {journal} {Electron. Mater.}\ }\textbf {\bibinfo {volume} {7}},\
  \bibinfo {pages} {3} (\bibinfo {year} {2026})}\BibitemShut {NoStop}%
\bibitem [{\citenamefont {Maass}\ \emph {et~al.}(1992)\citenamefont {Maass},
  \citenamefont {Bunde},\ and\ \citenamefont {Ingram}}]{Maass/etal:1992}%
  \BibitemOpen
  \bibfield  {author} {\bibinfo {author} {\bibfnamefont {P.}~\bibnamefont
  {Maass}}, \bibinfo {author} {\bibfnamefont {A.}~\bibnamefont {Bunde}},\ and\
  \bibinfo {author} {\bibfnamefont {M.~D.}\ \bibnamefont {Ingram}},\ }\bibfield
   {title} {\bibinfo {title} {Ion transport anomalies in glasses},\ }\href
  {https://doi.org/10.1103/PhysRevLett.68.3064} {\bibfield  {journal} {\bibinfo
   {journal} {Phys. Rev. Lett.}\ }\textbf {\bibinfo {volume} {68}},\ \bibinfo
  {pages} {3064} (\bibinfo {year} {1992})}\BibitemShut {NoStop}%
\bibitem [{\citenamefont {Hunt}(1999)}]{Hunt:1999}%
  \BibitemOpen
  \bibfield  {author} {\bibinfo {author} {\bibfnamefont {A.~G.}\ \bibnamefont
  {Hunt}},\ }\bibfield  {title} {\bibinfo {title} {Mixed-alkali effect: some
  new results},\ }\href
  {https://doi.org/https://doi.org/10.1016/S0022-3093(99)00423-8} {\bibfield
  {journal} {\bibinfo  {journal} {J. Non-Cryst. Solids}\ }\textbf {\bibinfo
  {volume} {255}},\ \bibinfo {pages} {47} (\bibinfo {year} {1999})}\BibitemShut
  {NoStop}%
\bibitem [{\citenamefont {Baranovskii}\ and\ \citenamefont
  {Cordes}(1999)}]{Baranovskii/Cordes:1999}%
  \BibitemOpen
  \bibfield  {author} {\bibinfo {author} {\bibfnamefont {S.~D.}\ \bibnamefont
  {Baranovskii}}\ and\ \bibinfo {author} {\bibfnamefont {H.}~\bibnamefont
  {Cordes}},\ }\bibfield  {title} {\bibinfo {title} {On the conduction
  mechanism in ionic glasses},\ }\href {https://doi.org/10.1063/1.480081}
  {\bibfield  {journal} {\bibinfo  {journal} {J. Chem. Phys.}\ }\textbf
  {\bibinfo {volume} {111}},\ \bibinfo {pages} {7546} (\bibinfo {year}
  {1999})}\BibitemShut {NoStop}%
\bibitem [{\citenamefont {Maass}(1999)}]{Maass:1999}%
  \BibitemOpen
  \bibfield  {author} {\bibinfo {author} {\bibfnamefont {P.}~\bibnamefont
  {Maass}},\ }\bibfield  {title} {\bibinfo {title} {Towards a theory for the
  mixed alkali effect in glasses},\ }\href
  {https://doi.org/https://doi.org/10.1016/S0022-3093(99)00422-6} {\bibfield
  {journal} {\bibinfo  {journal} {J. Non-Cryst. Solids}\ }\textbf {\bibinfo
  {volume} {255}},\ \bibinfo {pages} {35} (\bibinfo {year} {1999})}\BibitemShut
  {NoStop}%
\bibitem [{\citenamefont {Kirchheim}(2000)}]{Kirchheim:2000}%
  \BibitemOpen
  \bibfield  {author} {\bibinfo {author} {\bibfnamefont {R.}~\bibnamefont
  {Kirchheim}},\ }\bibfield  {title} {\bibinfo {title} {The mixed alkali effect
  as a consequence of network density and site energy distribution},\ }\href
  {https://doi.org/https://doi.org/10.1016/S0022-3093(00)00165-4} {\bibfield
  {journal} {\bibinfo  {journal} {J. Non-Cryst. Solids}\ }\textbf {\bibinfo
  {volume} {272}},\ \bibinfo {pages} {85 } (\bibinfo {year}
  {2000})}\BibitemShut {NoStop}%
\bibitem [{\citenamefont {Greaves}(1998)}]{Greaves:1998}%
  \BibitemOpen
  \bibfield  {author} {\bibinfo {author} {\bibfnamefont {G.~N.}\ \bibnamefont
  {Greaves}},\ }\bibfield  {title} {\bibinfo {title} {Structural studies of the
  mixed alkali effect in disilicate glasses},\ }\href
  {https://doi.org/https://doi.org/10.1016/S0167-2738(97)00471-2} {\bibfield
  {journal} {\bibinfo  {journal} {Solid State Ionics}\ }\textbf {\bibinfo
  {volume} {105}},\ \bibinfo {pages} {243} (\bibinfo {year}
  {1998})}\BibitemShut {NoStop}%
\bibitem [{\citenamefont {Tsuchida}\ \emph {et~al.}(2012)\citenamefont
  {Tsuchida}, \citenamefont {Schneider}, \citenamefont {Deshpande},\ and\
  \citenamefont {Eckert}}]{Tsuchida/etal:2012}%
  \BibitemOpen
  \bibfield  {author} {\bibinfo {author} {\bibfnamefont {J.}~\bibnamefont
  {Tsuchida}}, \bibinfo {author} {\bibfnamefont {J.}~\bibnamefont {Schneider}},
  \bibinfo {author} {\bibfnamefont {R.~R.}\ \bibnamefont {Deshpande}},\ and\
  \bibinfo {author} {\bibfnamefont {H.}~\bibnamefont {Eckert}},\ }\bibfield
  {title} {\bibinfo {title} {Cation distribution and local order in mixed
  sodium metaphosphate glasses},\ }\href {https://doi.org/10.1021/jp308268h}
  {\bibfield  {journal} {\bibinfo  {journal} {J. Phys. Chem. C}\ }\textbf
  {\bibinfo {volume} {116}},\ \bibinfo {pages} {24449} (\bibinfo {year}
  {2012})}\BibitemShut {NoStop}%
\bibitem [{\citenamefont {Nogueira~dos Santos}\ \emph
  {et~al.}(2025)\citenamefont {Nogueira~dos Santos}, \citenamefont
  {de~Melo~Marques}, \citenamefont {Nieto~Mun{\~o}z}, \citenamefont
  {Martins~Rodrigues}, \citenamefont {Schneider},\ and\ \citenamefont
  {Esquina~Tsuchida}}]{NogueiradosSantos/etal:2025}%
  \BibitemOpen
  \bibfield  {author} {\bibinfo {author} {\bibfnamefont {I.~M.}\ \bibnamefont
  {Nogueira~dos Santos}}, \bibinfo {author} {\bibfnamefont {F.~A.}\
  \bibnamefont {de~Melo~Marques}}, \bibinfo {author} {\bibfnamefont {A.~M.}\
  \bibnamefont {Nieto~Mun{\~o}z}}, \bibinfo {author} {\bibfnamefont {A.~C.}\
  \bibnamefont {Martins~Rodrigues}}, \bibinfo {author} {\bibfnamefont {J.~F.}\
  \bibnamefont {Schneider}},\ and\ \bibinfo {author} {\bibfnamefont
  {J.}~\bibnamefont {Esquina~Tsuchida}},\ }\bibfield  {title} {\bibinfo {title}
  {Structural characterization and study of the mixed-ion effect in {K-Li}
  metaphosphate glasses},\ }\href {https://doi.org/10.1021/acsomega.5c01027}
  {\bibfield  {journal} {\bibinfo  {journal} {ACS Omega}\ }\textbf {\bibinfo
  {volume} {10}},\ \bibinfo {pages} {15646} (\bibinfo {year}
  {2025})}\BibitemShut {NoStop}%
\bibitem [{\citenamefont {Habasaki}\ \emph {et~al.}(1995)\citenamefont
  {Habasaki}, \citenamefont {Okada},\ and\ \citenamefont
  {Hiwatari}}]{Habasaki/etal:1995}%
  \BibitemOpen
  \bibfield  {author} {\bibinfo {author} {\bibfnamefont {J.}~\bibnamefont
  {Habasaki}}, \bibinfo {author} {\bibfnamefont {I.}~\bibnamefont {Okada}},\
  and\ \bibinfo {author} {\bibfnamefont {Y.}~\bibnamefont {Hiwatari}},\
  }\bibfield  {title} {\bibinfo {title} {{MD} study of the mixed alkali effect
  in a lithium-potassium metasilicate glass},\ }\href
  {https://doi.org/https://doi.org/10.1016/0022-3093(94)00529-X} {\bibfield
  {journal} {\bibinfo  {journal} {J. Non-Cryst. Solids}\ }\textbf {\bibinfo
  {volume} {183}},\ \bibinfo {pages} {12} (\bibinfo {year} {1995})}\BibitemShut
  {NoStop}%
\bibitem [{\citenamefont {Habasaki}\ \emph {et~al.}(2004)\citenamefont
  {Habasaki}, \citenamefont {Ngai},\ and\ \citenamefont
  {Hiwatari}}]{Habasaki/etal:2004}%
  \BibitemOpen
  \bibfield  {author} {\bibinfo {author} {\bibfnamefont {J.}~\bibnamefont
  {Habasaki}}, \bibinfo {author} {\bibfnamefont {K.~L.}\ \bibnamefont {Ngai}},\
  and\ \bibinfo {author} {\bibfnamefont {Y.}~\bibnamefont {Hiwatari}},\
  }\bibfield  {title} {\bibinfo {title} {``{C}ooperativity blockage'' in the
  mixed alkali effect as revealed by molecular-dynamics simulations of alkali
  metasilicate glass},\ }\href {https://doi.org/10.1063/1.1759614} {\bibfield
  {journal} {\bibinfo  {journal} {J. Chem. Phys.}\ }\textbf {\bibinfo {volume}
  {121}},\ \bibinfo {pages} {925} (\bibinfo {year} {2004})}\BibitemShut
  {NoStop}%
\bibitem [{\citenamefont {Balbuena}\ \emph {et~al.}(2014)\citenamefont
  {Balbuena}, \citenamefont {Frechero},\ and\ \citenamefont
  {Montani}}]{Balbuena/etal:2014}%
  \BibitemOpen
  \bibfield  {author} {\bibinfo {author} {\bibfnamefont {C.}~\bibnamefont
  {Balbuena}}, \bibinfo {author} {\bibfnamefont {M.~A.}\ \bibnamefont
  {Frechero}},\ and\ \bibinfo {author} {\bibfnamefont {R.~A.}\ \bibnamefont
  {Montani}},\ }\bibfield  {title} {\bibinfo {title} {Channel diffusion in a
  lithium--potassium metasilicate glass using the isoconfigurational ensemble:
  Towards a scenario for the mixed alkali effect},\ }\href
  {https://doi.org/https://doi.org/10.1016/j.jnoncrysol.2014.09.001} {\bibfield
   {journal} {\bibinfo  {journal} {J. Non-Cryst. Solids}\ }\textbf {\bibinfo
  {volume} {405}},\ \bibinfo {pages} {124} (\bibinfo {year}
  {2014})}\BibitemShut {NoStop}%
\bibitem [{\citenamefont {Lodesani}\ \emph {et~al.}(2020)\citenamefont
  {Lodesani}, \citenamefont {Menziani}, \citenamefont {Hijiya}, \citenamefont
  {Takato}, \citenamefont {Urata},\ and\ \citenamefont
  {Pedone}}]{Lodesani/etal:2020}%
  \BibitemOpen
  \bibfield  {author} {\bibinfo {author} {\bibfnamefont {F.}~\bibnamefont
  {Lodesani}}, \bibinfo {author} {\bibfnamefont {M.~C.}\ \bibnamefont
  {Menziani}}, \bibinfo {author} {\bibfnamefont {H.}~\bibnamefont {Hijiya}},
  \bibinfo {author} {\bibfnamefont {Y.}~\bibnamefont {Takato}}, \bibinfo
  {author} {\bibfnamefont {S.}~\bibnamefont {Urata}},\ and\ \bibinfo {author}
  {\bibfnamefont {A.}~\bibnamefont {Pedone}},\ }\bibfield  {title} {\bibinfo
  {title} {Structural origins of the mixed alkali effect in alkali
  aluminosilicate glasses: Molecular dynamics study and its assessment},\
  }\href {https://doi.org/10.1038/s41598-020-59875-7} {\bibfield  {journal}
  {\bibinfo  {journal} {Sci. Rep.}\ }\textbf {\bibinfo {volume} {10}},\
  \bibinfo {pages} {2906} (\bibinfo {year} {2020})}\BibitemShut {NoStop}%
\bibitem [{\citenamefont {Noritake}\ and\ \citenamefont
  {Naito}(2023)}]{Noritake/Naito:2023}%
  \BibitemOpen
  \bibfield  {author} {\bibinfo {author} {\bibfnamefont {F.}~\bibnamefont
  {Noritake}}\ and\ \bibinfo {author} {\bibfnamefont {S.}~\bibnamefont
  {Naito}},\ }\bibfield  {title} {\bibinfo {title} {Mechanism of mixed alkali
  effect in silicate glass/liquid: {P}athway and network analysis},\ }\href
  {https://doi.org/https://doi.org/10.1016/j.jnoncrysol.2023.122321} {\bibfield
   {journal} {\bibinfo  {journal} {J. Non-Cryst. Solids}\ }\textbf {\bibinfo
  {volume} {610}},\ \bibinfo {pages} {122321} (\bibinfo {year}
  {2023})}\BibitemShut {NoStop}%
\bibitem [{\citenamefont {Lammert}\ and\ \citenamefont
  {Heuer}(2005)}]{Lammert/Heuer:2005}%
  \BibitemOpen
  \bibfield  {author} {\bibinfo {author} {\bibfnamefont {H.}~\bibnamefont
  {Lammert}}\ and\ \bibinfo {author} {\bibfnamefont {A.}~\bibnamefont
  {Heuer}},\ }\bibfield  {title} {\bibinfo {title} {Contributions to the
  mixed-alkali effect in molecular dynamics simulations of alkali silicate
  glasses},\ }\href {https://doi.org/10.1103/PhysRevB.72.214202} {\bibfield
  {journal} {\bibinfo  {journal} {Phys. Rev. B}\ }\textbf {\bibinfo {volume}
  {72}},\ \bibinfo {pages} {214202} (\bibinfo {year} {2005})}\BibitemShut
  {NoStop}%
\bibitem [{\citenamefont {Leiber}\ \emph {et~al.}(2026)\citenamefont {Leiber},
  \citenamefont {Fischer}, \citenamefont {Lohmann},\ and\ \citenamefont
  {Maass}}]{Leiber/etal:2026}%
  \BibitemOpen
  \bibfield  {author} {\bibinfo {author} {\bibfnamefont {J.}~\bibnamefont
  {Leiber}}, \bibinfo {author} {\bibfnamefont {Q.~E.}\ \bibnamefont {Fischer}},
  \bibinfo {author} {\bibfnamefont {S.}~\bibnamefont {Lohmann}},\ and\ \bibinfo
  {author} {\bibfnamefont {P.}~\bibnamefont {Maass}},\ }\bibfield  {title}
  {\bibinfo {title} {Theory for the mixed alkali effect in glasses},\ }\href
  {https://doi.org/10.1063/5.0343588} {\bibfield  {journal} {\bibinfo
  {journal} {J. Chem. Phys.}\ }\textbf {\bibinfo {volume} {165}},\ \bibinfo
  {pages} {064504} (\bibinfo {year} {2026})}\BibitemShut {NoStop}%
\bibitem [{\citenamefont {Miller}\ and\ \citenamefont
  {Abrahams}(1960)}]{Miller/Abrahams:1960}%
  \BibitemOpen
  \bibfield  {author} {\bibinfo {author} {\bibfnamefont {A.}~\bibnamefont
  {Miller}}\ and\ \bibinfo {author} {\bibfnamefont {E.}~\bibnamefont
  {Abrahams}},\ }\bibfield  {title} {\bibinfo {title} {Impurity conduction at
  low concentrations},\ }\href {https://doi.org/10.1103/PhysRev.120.745}
  {\bibfield  {journal} {\bibinfo  {journal} {Phys. Rev.}\ }\textbf {\bibinfo
  {volume} {120}},\ \bibinfo {pages} {745} (\bibinfo {year}
  {1960})}\BibitemShut {NoStop}%
\bibitem [{\citenamefont {Ambegaokar}\ \emph {et~al.}(1971)\citenamefont
  {Ambegaokar}, \citenamefont {Halperin},\ and\ \citenamefont
  {Langer}}]{Ambegaokar/etal:1971}%
  \BibitemOpen
  \bibfield  {author} {\bibinfo {author} {\bibfnamefont {V.}~\bibnamefont
  {Ambegaokar}}, \bibinfo {author} {\bibfnamefont {B.~I.}\ \bibnamefont
  {Halperin}},\ and\ \bibinfo {author} {\bibfnamefont {J.~S.}\ \bibnamefont
  {Langer}},\ }\bibfield  {title} {\bibinfo {title} {Hopping conductivity in
  disordered systems},\ }\href {https://doi.org/10.1103/PhysRevB.4.2612}
  {\bibfield  {journal} {\bibinfo  {journal} {Phys. Rev. B}\ }\textbf {\bibinfo
  {volume} {4}},\ \bibinfo {pages} {2612} (\bibinfo {year} {1971})}\BibitemShut
  {NoStop}%
\bibitem [{\citenamefont {Peibst}\ \emph {et~al.}(2005)\citenamefont {Peibst},
  \citenamefont {Schott},\ and\ \citenamefont {Maass}}]{Peibst/etal:2005}%
  \BibitemOpen
  \bibfield  {author} {\bibinfo {author} {\bibfnamefont {R.}~\bibnamefont
  {Peibst}}, \bibinfo {author} {\bibfnamefont {S.}~\bibnamefont {Schott}},\
  and\ \bibinfo {author} {\bibfnamefont {P.}~\bibnamefont {Maass}},\ }\bibfield
   {title} {\bibinfo {title} {Internal friction and vulnerability of mixed
  alkali glasses},\ }\href {https://doi.org/10.1103/PhysRevLett.95.115901}
  {\bibfield  {journal} {\bibinfo  {journal} {Phys. Rev. Lett.}\ }\textbf
  {\bibinfo {volume} {95}},\ \bibinfo {pages} {115901} (\bibinfo {year}
  {2005})}\BibitemShut {NoStop}%
\bibitem [{\citenamefont {Shelby}(1965)}]{Shelby:1965}%
  \BibitemOpen
  \bibfield  {author} {\bibinfo {author} {\bibfnamefont {J.~E.}\ \bibnamefont
  {Shelby}},\ }\emph {\bibinfo {title} {Internal friction of mixed-alkali
  silicate glasses}},\ \href
  {https://scholarsmine.mst.edu/doctoral_dissertations/1867/} {Ph.D. thesis},\
  \bibinfo  {school} {Missouri University of Science and Technology} (\bibinfo
  {year} {1965})\BibitemShut {NoStop}%
\bibitem [{\citenamefont {Maass}\ \emph {et~al.}(1995)\citenamefont {Maass},
  \citenamefont {Meyer},\ and\ \citenamefont {Bunde}}]{Maass/etal:1995}%
  \BibitemOpen
  \bibfield  {author} {\bibinfo {author} {\bibfnamefont {P.}~\bibnamefont
  {Maass}}, \bibinfo {author} {\bibfnamefont {M.}~\bibnamefont {Meyer}},\ and\
  \bibinfo {author} {\bibfnamefont {A.}~\bibnamefont {Bunde}},\ }\bibfield
  {title} {\bibinfo {title} {Nonstandard relaxation behavior in ionically
  conducting materials},\ }\href {https://doi.org/10.1103/PhysRevB.51.8164}
  {\bibfield  {journal} {\bibinfo  {journal} {Phys. Rev. B}\ }\textbf {\bibinfo
  {volume} {51}},\ \bibinfo {pages} {8164} (\bibinfo {year}
  {1995})}\BibitemShut {NoStop}%
\bibitem [{\citenamefont {Lohmann}\ \emph {et~al.}(2026)\citenamefont
  {Lohmann}, \citenamefont {Fischer}, \citenamefont {Leiber},\ and\
  \citenamefont {Maass}}]{Lohmann/etal:2026}%
  \BibitemOpen
  \bibfield  {author} {\bibinfo {author} {\bibfnamefont {S.}~\bibnamefont
  {Lohmann}}, \bibinfo {author} {\bibfnamefont {Q.~E.}\ \bibnamefont
  {Fischer}}, \bibinfo {author} {\bibfnamefont {J.}~\bibnamefont {Leiber}},\
  and\ \bibinfo {author} {\bibfnamefont {P.}~\bibnamefont {Maass}},\ }\bibfield
   {title} {\bibinfo {title} {Universal time-temperature scaling of
  conductivities in random site energy and associated random barrier models},\
  }\href {https://doi.org/10.1103/mvpq-96kr} {\bibfield  {journal} {\bibinfo
  {journal} {Phys. Rev. B}\ }\textbf {\bibinfo {volume} {113}},\ \bibinfo
  {pages} {184201} (\bibinfo {year} {2026})}\BibitemShut {NoStop}%
\bibitem [{\citenamefont {Schr{\o}der}(2000)}]{Schroeder:2000}%
  \BibitemOpen
  \bibfield  {author} {\bibinfo {author} {\bibfnamefont {T.~B.}\ \bibnamefont
  {Schr{\o}der}},\ }\emph {\bibinfo {title} {Hopping in {D}isordered {M}edia:
  {A} {M}odel {G}lass {F}ormer and a {H}opping {M}odel}},\ \href@noop {} {Ph.D.
  thesis},\ \bibinfo  {school} {Roskilde Universitet, Denmark} (\bibinfo {year}
  {2000}),\ \bibinfo {note} {arXiv:cond-mat/0005127}\BibitemShut {NoStop}%
\bibitem [{\citenamefont {Schr{\o}der}(2008)}]{Schroeder:2008}%
  \BibitemOpen
  \bibfield  {author} {\bibinfo {author} {\bibfnamefont {T.~B.}\ \bibnamefont
  {Schr{\o}der}},\ }\bibfield  {title} {\bibinfo {title} {Efficient numerical
  method for studying random walks on disordered structures},\ }\href
  {https://doi.org/10.1209/0295-5075/81/30002} {\bibfield  {journal} {\bibinfo
  {journal} {Europhys. Lett.}\ }\textbf {\bibinfo {volume} {81}},\ \bibinfo
  {pages} {30002} (\bibinfo {year} {2008})}\BibitemShut {NoStop}%
\bibitem [{\citenamefont {Bosi}\ and\ \citenamefont
  {Maass}(2022)}]{Bosi/Maass:2022}%
  \BibitemOpen
  \bibfield  {author} {\bibinfo {author} {\bibfnamefont {M.}~\bibnamefont
  {Bosi}}\ and\ \bibinfo {author} {\bibfnamefont {P.}~\bibnamefont {Maass}},\
  }\bibfield  {title} {\bibinfo {title} {Predicting conductivities of alkali
  borophosphate glasses based on site energy distributions derived from network
  former unit concentrations},\ }\href
  {https://doi.org/doi:10.1515/zpch-2021-3137} {\bibfield  {journal} {\bibinfo
  {journal} {Z. Phys. Chem.}\ }\textbf {\bibinfo {volume} {236}},\ \bibinfo
  {pages} {1055} (\bibinfo {year} {2022})}\BibitemShut {NoStop}%
\bibitem [{\citenamefont {Bosi}\ \emph {et~al.}(2021)\citenamefont {Bosi},
  \citenamefont {Fischer},\ and\ \citenamefont {Maass}}]{Bosi/etal:2021}%
  \BibitemOpen
  \bibfield  {author} {\bibinfo {author} {\bibfnamefont {M.}~\bibnamefont
  {Bosi}}, \bibinfo {author} {\bibfnamefont {J.}~\bibnamefont {Fischer}},\ and\
  \bibinfo {author} {\bibfnamefont {P.}~\bibnamefont {Maass}},\ }\bibfield
  {title} {\bibinfo {title} {Network-forming units, energy landscapes, and
  conductivity activation energies in alkali borophosphate glasses: Analytical
  approaches},\ }\href {https://doi.org/10.1021/acs.jpcc.0c09960} {\bibfield
  {journal} {\bibinfo  {journal} {J. Phys. Chem. C}\ }\textbf {\bibinfo
  {volume} {125}},\ \bibinfo {pages} {6260} (\bibinfo {year}
  {2021})}\BibitemShut {NoStop}%
\bibitem [{\citenamefont {Schuch}\ \emph {et~al.}(2012)\citenamefont {Schuch},
  \citenamefont {Christensen}, \citenamefont {Trott}, \citenamefont {Maass},\
  and\ \citenamefont {Martin}}]{Schuch/etal:2012}%
  \BibitemOpen
  \bibfield  {author} {\bibinfo {author} {\bibfnamefont {M.}~\bibnamefont
  {Schuch}}, \bibinfo {author} {\bibfnamefont {R.}~\bibnamefont {Christensen}},
  \bibinfo {author} {\bibfnamefont {C.}~\bibnamefont {Trott}}, \bibinfo
  {author} {\bibfnamefont {P.}~\bibnamefont {Maass}},\ and\ \bibinfo {author}
  {\bibfnamefont {S.~W.}\ \bibnamefont {Martin}},\ }\bibfield  {title}
  {\bibinfo {title} {Investigation of the structures of sodium borophosphate
  glasses by reverse {M}onte {C}arlo modeling to examine the origins of the
  mixed glass former effect},\ }\href {https://doi.org/10.1021/jp2085654}
  {\bibfield  {journal} {\bibinfo  {journal} {J. Phys. Chem. C}\ }\textbf
  {\bibinfo {volume} {116}},\ \bibinfo {pages} {1503} (\bibinfo {year}
  {2012})}\BibitemShut {NoStop}%
\bibitem [{\citenamefont {Karlsson}\ \emph {et~al.}(2015)\citenamefont
  {Karlsson}, \citenamefont {Schuch}, \citenamefont {Christensen},
  \citenamefont {Maass}, \citenamefont {Martin}, \citenamefont {Imberti},\ and\
  \citenamefont {Matic}}]{Karlsson/etal:2015}%
  \BibitemOpen
  \bibfield  {author} {\bibinfo {author} {\bibfnamefont {M.}~\bibnamefont
  {Karlsson}}, \bibinfo {author} {\bibfnamefont {M.}~\bibnamefont {Schuch}},
  \bibinfo {author} {\bibfnamefont {R.}~\bibnamefont {Christensen}}, \bibinfo
  {author} {\bibfnamefont {P.}~\bibnamefont {Maass}}, \bibinfo {author}
  {\bibfnamefont {S.~W.}\ \bibnamefont {Martin}}, \bibinfo {author}
  {\bibfnamefont {S.}~\bibnamefont {Imberti}},\ and\ \bibinfo {author}
  {\bibfnamefont {A.}~\bibnamefont {Matic}},\ }\bibfield  {title} {\bibinfo
  {title} {Structural origin of the mixed glass former effect in sodium
  borophosphate glasses investigated with neutron diffraction and reverse
  {M}onte {C}arlo modeling},\ }\href {https://doi.org/10.1021/acs.jpcc.5b09176}
  {\bibfield  {journal} {\bibinfo  {journal} {J. Phys. Chem. B}\ }\textbf
  {\bibinfo {volume} {119}},\ \bibinfo {pages} {27275} (\bibinfo {year}
  {2015})}\BibitemShut {NoStop}%
\bibitem [{\citenamefont {Kulkarni}\ \emph {et~al.}(1998)\citenamefont
  {Kulkarni}, \citenamefont {Lunkenheimer},\ and\ \citenamefont
  {Loidl}}]{Kulkarni/etal:1998}%
  \BibitemOpen
  \bibfield  {author} {\bibinfo {author} {\bibfnamefont {A.~R.}\ \bibnamefont
  {Kulkarni}}, \bibinfo {author} {\bibfnamefont {P.}~\bibnamefont
  {Lunkenheimer}},\ and\ \bibinfo {author} {\bibfnamefont {A.}~\bibnamefont
  {Loidl}},\ }\bibfield  {title} {\bibinfo {title} {Scaling behaviour in the
  frequency dependent conductivity of mixed alkali glasses},\ }\href
  {https://doi.org/https://doi.org/10.1016/S0167-2738(98)00210-0} {\bibfield
  {journal} {\bibinfo  {journal} {Solid State Ionics}\ }\textbf {\bibinfo
  {volume} {112}},\ \bibinfo {pages} {69} (\bibinfo {year} {1998})}\BibitemShut
  {NoStop}%
\bibitem [{\citenamefont {Chen}\ \emph {et~al.}(1992)\citenamefont {Chen},
  \citenamefont {Yang}, \citenamefont {Durand}, \citenamefont {Pradel},\ and\
  \citenamefont {Ribes}}]{Chen/etal:1992}%
  \BibitemOpen
  \bibfield  {author} {\bibinfo {author} {\bibfnamefont {R.}~\bibnamefont
  {Chen}}, \bibinfo {author} {\bibfnamefont {R.}~\bibnamefont {Yang}}, \bibinfo
  {author} {\bibfnamefont {B.}~\bibnamefont {Durand}}, \bibinfo {author}
  {\bibfnamefont {A.}~\bibnamefont {Pradel}},\ and\ \bibinfo {author}
  {\bibfnamefont {M.}~\bibnamefont {Ribes}},\ }\bibfield  {title} {\bibinfo
  {title} {A study of the mixed alkali effect by frequency-dependent
  conductivity in {L}i$_2${O}-{N}a$_2${O}-{P}$_2${O}$_5$ glasses},\ }\href
  {https://doi.org/https://doi.org/10.1016/0167-2738(92)90312-D} {\bibfield
  {journal} {\bibinfo  {journal} {Solid State Ionics}\ }\textbf {\bibinfo
  {volume} {53-56}},\ \bibinfo {pages} {1194} (\bibinfo {year}
  {1992})}\BibitemShut {NoStop}%
\bibitem [{\citenamefont {Rathan}\ \emph {et~al.}(2022)\citenamefont {Rathan},
  \citenamefont {Murugaraj}, \citenamefont {Govindaraj}, \citenamefont
  {Aravinth},\ and\ \citenamefont {Ramasamy}}]{Rathan/etal:2022}%
  \BibitemOpen
  \bibfield  {author} {\bibinfo {author} {\bibfnamefont {S.~V.}\ \bibnamefont
  {Rathan}}, \bibinfo {author} {\bibfnamefont {R.}~\bibnamefont {Murugaraj}},
  \bibinfo {author} {\bibfnamefont {G.}~\bibnamefont {Govindaraj}}, \bibinfo
  {author} {\bibfnamefont {K.}~\bibnamefont {Aravinth}},\ and\ \bibinfo
  {author} {\bibfnamefont {P.}~\bibnamefont {Ramasamy}},\ }\bibfield  {title}
  {\bibinfo {title} {iniaturization of thermal, glass formation, and electrical
  properties with contrive scaling in mixed alkali
  ({L}i$_{1-x}${N}a$_x$)$_5${TiP}$_3${O}$_{12}$ ($0.0\le x\le 1.0$) phosphate
  glasses},\ }\href
  {https://doi.org/https://doi.org/10.1016/j.jnoncrysol.2022.121735} {\bibfield
   {journal} {\bibinfo  {journal} {J. Non-Cryst. Solids}\ }\textbf {\bibinfo
  {volume} {591}},\ \bibinfo {pages} {121735} (\bibinfo {year}
  {2022})}\BibitemShut {NoStop}%
\bibitem [{\citenamefont {Summerfield}(1985)}]{Summerfield:1985}%
  \BibitemOpen
  \bibfield  {author} {\bibinfo {author} {\bibfnamefont {S.}~\bibnamefont
  {Summerfield}},\ }\bibfield  {title} {\bibinfo {title} {Universal
  low-frequency behaviour in the a.c. hopping conductivity of disordered
  systems},\ }\href {https://doi.org/10.1080/13642818508243162} {\bibfield
  {journal} {\bibinfo  {journal} {Philos. Mag. B}\ }\textbf {\bibinfo {volume}
  {52}},\ \bibinfo {pages} {9} (\bibinfo {year} {1985})}\BibitemShut {NoStop}%
\bibitem [{\citenamefont {Roling}\ \emph {et~al.}(1997)\citenamefont {Roling},
  \citenamefont {Happe}, \citenamefont {Funke},\ and\ \citenamefont
  {Ingram}}]{Roling/etal:1997}%
  \BibitemOpen
  \bibfield  {author} {\bibinfo {author} {\bibfnamefont {B.}~\bibnamefont
  {Roling}}, \bibinfo {author} {\bibfnamefont {A.}~\bibnamefont {Happe}},
  \bibinfo {author} {\bibfnamefont {K.}~\bibnamefont {Funke}},\ and\ \bibinfo
  {author} {\bibfnamefont {M.~D.}\ \bibnamefont {Ingram}},\ }\bibfield  {title}
  {\bibinfo {title} {Carrier concentrations and relaxation spectroscopy: New
  information from scaling properties of conductivity spectra in ionically
  conducting glasses},\ }\href {https://doi.org/10.1103/PhysRevLett.78.2160}
  {\bibfield  {journal} {\bibinfo  {journal} {Phys. Rev. Lett.}\ }\textbf
  {\bibinfo {volume} {78}},\ \bibinfo {pages} {2160} (\bibinfo {year}
  {1997})}\BibitemShut {NoStop}%
\bibitem [{\citenamefont {Sidebottom}(1999)}]{Sidebottom:1999}%
  \BibitemOpen
  \bibfield  {author} {\bibinfo {author} {\bibfnamefont {D.~L.}\ \bibnamefont
  {Sidebottom}},\ }\bibfield  {title} {\bibinfo {title} {Universal approach for
  scaling the ac conductivity in ionic glasses},\ }\href
  {https://doi.org/10.1103/PhysRevLett.82.3653} {\bibfield  {journal} {\bibinfo
   {journal} {Phys. Rev. Lett.}\ }\textbf {\bibinfo {volume} {82}},\ \bibinfo
  {pages} {3653} (\bibinfo {year} {1999})}\BibitemShut {NoStop}%
\bibitem [{\citenamefont {Ghosh}\ and\ \citenamefont
  {Pan}(2000)}]{Ghosh/Pan:2000}%
  \BibitemOpen
  \bibfield  {author} {\bibinfo {author} {\bibfnamefont {A.}~\bibnamefont
  {Ghosh}}\ and\ \bibinfo {author} {\bibfnamefont {A.}~\bibnamefont {Pan}},\
  }\bibfield  {title} {\bibinfo {title} {Scaling of the conductivity spectra in
  ionic glasses: {D}ependence on the structure},\ }\href
  {https://doi.org/10.1103/PhysRevLett.84.2188} {\bibfield  {journal} {\bibinfo
   {journal} {Phys. Rev. Lett.}\ }\textbf {\bibinfo {volume} {84}},\ \bibinfo
  {pages} {2188} (\bibinfo {year} {2000})}\BibitemShut {NoStop}%
\bibitem [{\citenamefont {Dyre}\ and\ \citenamefont
  {Schr{\o}der}(2000)}]{Dyre/Schroeder:2000}%
  \BibitemOpen
  \bibfield  {author} {\bibinfo {author} {\bibfnamefont {J.~C.}\ \bibnamefont
  {Dyre}}\ and\ \bibinfo {author} {\bibfnamefont {T.~B.}\ \bibnamefont
  {Schr{\o}der}},\ }\bibfield  {title} {\bibinfo {title} {Universality of ac
  conduction in disordered solids},\ }\href
  {https://doi.org/10.1103/RevModPhys.72.873} {\bibfield  {journal} {\bibinfo
  {journal} {Rev. Mod. Phys.}\ }\textbf {\bibinfo {volume} {72}},\ \bibinfo
  {pages} {873} (\bibinfo {year} {2000})}\BibitemShut {NoStop}%
\bibitem [{\citenamefont {Schr\o{}der}\ and\ \citenamefont
  {Dyre}(2000)}]{Schroeder/Dyre:2000}%
  \BibitemOpen
  \bibfield  {author} {\bibinfo {author} {\bibfnamefont {T.~B.}\ \bibnamefont
  {Schr\o{}der}}\ and\ \bibinfo {author} {\bibfnamefont {J.~C.}\ \bibnamefont
  {Dyre}},\ }\bibfield  {title} {\bibinfo {title} {Scaling and universality of
  ac conduction in disordered solids},\ }\href
  {https://doi.org/10.1103/PhysRevLett.84.310} {\bibfield  {journal} {\bibinfo
  {journal} {Phys. Rev. Lett.}\ }\textbf {\bibinfo {volume} {84}},\ \bibinfo
  {pages} {310} (\bibinfo {year} {2000})}\BibitemShut {NoStop}%
\bibitem [{\citenamefont {Schr\o{}der}\ and\ \citenamefont
  {Dyre}(2008)}]{Schroeder/Dyre:2008}%
  \BibitemOpen
  \bibfield  {author} {\bibinfo {author} {\bibfnamefont {T.~B.}\ \bibnamefont
  {Schr\o{}der}}\ and\ \bibinfo {author} {\bibfnamefont {J.~C.}\ \bibnamefont
  {Dyre}},\ }\bibfield  {title} {\bibinfo {title} {{ac} {H}opping conduction at
  extreme disorder takes place on the percolating cluster},\ }\href
  {https://doi.org/10.1103/PhysRevLett.101.025901} {\bibfield  {journal}
  {\bibinfo  {journal} {Phys. Rev. Lett.}\ }\textbf {\bibinfo {volume} {101}},\
  \bibinfo {pages} {025901} (\bibinfo {year} {2008})}\BibitemShut {NoStop}%
\bibitem [{\citenamefont {Shelby~JR.}\ and\ \citenamefont
  {Day}(1969)}]{Shelby/Day:1969}%
  \BibitemOpen
  \bibfield  {author} {\bibinfo {author} {\bibfnamefont {J.~E.}\ \bibnamefont
  {Shelby~JR.}}\ and\ \bibinfo {author} {\bibfnamefont {D.~E.}\ \bibnamefont
  {Day}},\ }\bibfield  {title} {\bibinfo {title} {Mechanical relaxations in
  mixed-alkali silicate glasses: I, {R}esults},\ }\href
  {https://doi.org/https://doi.org/10.1111/j.1151-2916.1969.tb13358.x}
  {\bibfield  {journal} {\bibinfo  {journal} {J. Am. Ceram. Soc.}\ }\textbf
  {\bibinfo {volume} {52}},\ \bibinfo {pages} {169} (\bibinfo {year}
  {1969})}\BibitemShut {NoStop}%
\bibitem [{\citenamefont {Shelby~JR.}\ and\ \citenamefont
  {Day}(1970)}]{Shelby/Day:1970}%
  \BibitemOpen
  \bibfield  {author} {\bibinfo {author} {\bibfnamefont {J.~E.}\ \bibnamefont
  {Shelby~JR.}}\ and\ \bibinfo {author} {\bibfnamefont {D.~E.}\ \bibnamefont
  {Day}},\ }\bibfield  {title} {\bibinfo {title} {Mechanical relaxations in
  mixed alkali silicate glasses: {II}, {D}iscussion},\ }\href
  {https://doi.org/https://doi.org/10.1111/j.1151-2916.1970.tb12067.x}
  {\bibfield  {journal} {\bibinfo  {journal} {J. Am. Ceram. Soc.}\ }\textbf
  {\bibinfo {volume} {53}},\ \bibinfo {pages} {182} (\bibinfo {year}
  {1970})}\BibitemShut {NoStop}%
\bibitem [{\citenamefont {McVay}\ and\ \citenamefont
  {Day}(1970)}]{McVay/Day:1970}%
  \BibitemOpen
  \bibfield  {author} {\bibinfo {author} {\bibfnamefont {G.~L.}\ \bibnamefont
  {McVay}}\ and\ \bibinfo {author} {\bibfnamefont {D.~E.}\ \bibnamefont
  {Day}},\ }\bibfield  {title} {\bibinfo {title} {Diffusion and internal
  friction in {N}a-{R}b silicate glasses},\ }\href
  {https://doi.org/https://doi.org/10.1111/j.1151-2916.1970.tb16002.x}
  {\bibfield  {journal} {\bibinfo  {journal} {J. Am. Ceram. Soc.}\ }\textbf
  {\bibinfo {volume} {53}},\ \bibinfo {pages} {508} (\bibinfo {year}
  {1970})}\BibitemShut {NoStop}%
\bibitem [{\citenamefont {Sakamura}\ \emph {et~al.}(2003)\citenamefont
  {Sakamura}, \citenamefont {Suzuki}, \citenamefont {Miyazaki}, \citenamefont
  {Nakazawa},\ and\ \citenamefont {Katayama}}]{Sakamura/etal:2003}%
  \BibitemOpen
  \bibfield  {author} {\bibinfo {author} {\bibfnamefont {M.}~\bibnamefont
  {Sakamura}}, \bibinfo {author} {\bibfnamefont {S.}~\bibnamefont {Suzuki}},
  \bibinfo {author} {\bibfnamefont {H.}~\bibnamefont {Miyazaki}}, \bibinfo
  {author} {\bibfnamefont {K.}~\bibnamefont {Nakazawa}},\ and\ \bibinfo
  {author} {\bibfnamefont {K.}~\bibnamefont {Katayama}},\ }\bibfield  {title}
  {\bibinfo {title} {Mixed alkali effect in internal friction of
  aluminogermanate glasses},\ }\href
  {https://doi.org/10.14853/pcersj.2003F.0.866.0} {\bibfield  {journal}
  {\bibinfo  {journal} {Preprints of Fall Meeting of The Ceramic Society of
  Japan}\ }\textbf {\bibinfo {volume} {2003F}},\ \bibinfo {pages} {866}
  (\bibinfo {year} {2003})}\BibitemShut {NoStop}%
\bibitem [{\citenamefont {Chocinski-Arnault}\ and\ \citenamefont
  {Riviere}(2007)}]{Chocinski-Arnault/Riviere:2007}%
  \BibitemOpen
  \bibfield  {author} {\bibinfo {author} {\bibfnamefont {L.}~\bibnamefont
  {Chocinski-Arnault}}\ and\ \bibinfo {author} {\bibfnamefont {A.}~\bibnamefont
  {Riviere}},\ }\bibfield  {title} {\bibinfo {title} {Isothermal mechanical
  relaxations in single and mixed alkali silicate glasses},\ }\href
  {https://doi.org/10.1080/14786430601021629} {\bibfield  {journal} {\bibinfo
  {journal} {Phil. Mag.}\ }\textbf {\bibinfo {volume} {87}},\ \bibinfo {pages}
  {749} (\bibinfo {year} {2007})}\BibitemShut {NoStop}%
\bibitem [{\citenamefont {Rivi{\`e}re}\ and\ \citenamefont
  {Chocinski-Arnault}(2009)}]{Riviere/Chocinski-Arnault:2009}%
  \BibitemOpen
  \bibfield  {author} {\bibinfo {author} {\bibfnamefont {A.}~\bibnamefont
  {Rivi{\`e}re}}\ and\ \bibinfo {author} {\bibfnamefont {L.}~\bibnamefont
  {Chocinski-Arnault}},\ }\bibfield  {title} {\bibinfo {title} {Mechanical
  relaxations in single and mixed alkali silicate glasses},\ }\href
  {https://doi.org/https://doi.org/10.1016/j.msea.2008.09.101} {\bibfield
  {journal} {\bibinfo  {journal} {Mater. Sci. Eng. A}\ }\textbf {\bibinfo
  {volume} {521-522}},\ \bibinfo {pages} {272} (\bibinfo {year}
  {2009})}\BibitemShut {NoStop}%
\bibitem [{\citenamefont {Zdaniewski}\ \emph {et~al.}(1979)\citenamefont
  {Zdaniewski}, \citenamefont {Rindone},\ and\ \citenamefont
  {Day}}]{Zdaniewski/etal:1979}%
  \BibitemOpen
  \bibfield  {author} {\bibinfo {author} {\bibfnamefont {W.~A.}\ \bibnamefont
  {Zdaniewski}}, \bibinfo {author} {\bibfnamefont {G.~E.}\ \bibnamefont
  {Rindone}},\ and\ \bibinfo {author} {\bibfnamefont {D.~E.}\ \bibnamefont
  {Day}},\ }\bibfield  {title} {\bibinfo {title} {The internal friction of
  glasses},\ }\href {https://doi.org/10.1007/BF00550707} {\bibfield  {journal}
  {\bibinfo  {journal} {J. Mater. Sci.}\ }\textbf {\bibinfo {volume} {14}},\
  \bibinfo {pages} {763} (\bibinfo {year} {1979})}\BibitemShut {NoStop}%
\bibitem [{\citenamefont {Maass}\ and\ \citenamefont
  {Peibst}(2006)}]{Maass/Peibst:2006}%
  \BibitemOpen
  \bibfield  {author} {\bibinfo {author} {\bibfnamefont {P.}~\bibnamefont
  {Maass}}\ and\ \bibinfo {author} {\bibfnamefont {R.}~\bibnamefont {Peibst}},\
  }\bibfield  {title} {\bibinfo {title} {Ion diffusion and mechanical losses in
  mixed alkali glasses},\ }\href
  {https://doi.org/https://doi.org/10.1016/j.jnoncrysol.2005.12.061} {\bibfield
   {journal} {\bibinfo  {journal} {J. Non-Cryst. Solids}\ }\textbf {\bibinfo
  {volume} {352}},\ \bibinfo {pages} {5178} (\bibinfo {year}
  {2006})}\BibitemShut {NoStop}%
\end{thebibliography}

%

\end{document}